\documentclass[11pt,tightenlines,superscriptaddress,floatfixolumn,english,aps,
notitlepage,nofootinbib]{revtex4-1}

\usepackage{amsmath}
\usepackage{amssymb}
\usepackage{esint}
\usepackage{graphicx}
\usepackage{hyperref}
\usepackage{color}
\usepackage{xcolor}
\usepackage[utf8]{inputenc} 
\usepackage{enumitem}
\usepackage{float}
\usepackage{array}
\usepackage{subfig}
\usepackage{soul}
\usepackage[makeroom]{cancel}

\usepackage{mathtools, nccmath}
\usepackage{cleveref}
\usepackage{amsfonts}

\crefrangeformat{equation}{Eqs. #3(#1)#4--#5(#2)#6}

\newcommand{\be}{\beta}

\begin{document}

\title{
Factorial cumulants from global baryon number conservation
}

\author{Micha{\l} Barej}
\email{michal.barej@fis.agh.edu.pl}
\affiliation{AGH University of Science and Technology,
Faculty of Physics and Applied Computer Science,
30-059 Krak\'ow, Poland}

\author{Adam Bzdak}
\email{adam.bzdak@fis.agh.edu.pl}
\affiliation{AGH University of Science and Technology,
Faculty of Physics and Applied Computer Science,
30-059 Krak\'ow, Poland}

\begin{abstract}
The proton, antiproton and mixed proton-antiproton factorial cumulants originating from the global conservation of baryon number are calculated analytically up to the sixth order. 
Our results can be directly tested in experiments.

\end{abstract}

\maketitle

\section{Introduction}
Many effective models of quantum chromodynamics (QCD) predict the first-order phase transition and the associated critical end point between the hadronic matter and quark-gluon plasma
\cite{Stephanov:2004wx,BraunMunzinger:2008tz,Braun-Munzinger:2015hba,Bzdak:2019pkr}. 
One of the main approaches to search for such structures in the QCD phase diagram is based on the investigation of fluctuations of, e.g., net-baryon number, net-charge or net-strangeness number 
\cite{Jeon:2000wg,Asakawa:2000wh,Gazdzicki:2003bb,Gorenstein:2003hk,Stephanov:2004wx,Koch:2005vg,Stephanov:2008qz,Cheng:2008zh,Fu:2009wy,Skokov:2010uh,Stephanov:2011pb,Karsch:2011gg,Schaefer:2011ex,Chen:2011am,Luo:2011rg,Zhou:2012ay,Wang:2012jr,Herold:2016uvv,Luo:2017faz,Szymanski:2019yho,Ratti:2019tvj} measured in relativistic heavy ion collisions, see also a recent review in Ref. \cite{Bzdak:2019pkr}. 

Higher-order cumulants, $\kappa_n$, of the multiplicity distribution can be used to quantify the properties of such fluctuations since they are proportional to the higher powers of the correlation length \cite{Stephanov:2008qz}. However, the cumulants mix the correlation functions of different orders, and thus in experimental situations might be challenging to interpret. Also, in practice, the cumulants might be dominated by the trivial term representing the average number of particles. 
To avoid these difficulties, the factorial cumulants, $\hat{C}_n$,\footnote{In this paper we adopt the notation of Ref. \cite{Bzdak:2019pkr}.} can be used as they represent the integrated genuine multi-particle correlation functions \cite{Botet:2002gj,Ling:2015yau,Bzdak:2016sxg,Bzdak:2019pkr}.

The factorial cumulants have already been successfully applied 
to the STAR data on net-proton fluctuations \cite{Adamczyk:2013dal,Luo:2015ewa,Adam:2020unf}, which unveiled rather unexpected source of strong three- and four-proton correlations in central Au+Au collisions at $\sqrt{s_{_{NN}} }=7.7~\mathrm{GeV}$ \cite{Bzdak:2016sxg}. It was later found that these correlations are consistent with a two-component (bimodal) proton multiplicity distribution \cite{Bzdak:2018uhv,Bzdak:2018axe}, which might indicate an interesting physics or a potential issue with the experimental data.

It is known that fluctuations and correlations related to the first-order phase transition or the critical end point may be misinterpreted because of the potentially significant contributions from various effects, which in this case play a role of the background. 
For instance, small fluctuations of the impact parameter and thus the number of wounded nucleons \cite{Bialas:1976ed} were studied, e.g., in Refs. \cite{Skokov:2012ds,Braun-Munzinger:2016yjz,Bzdak:2016jxo}. This effect may lead to significant corrections, as recently shown in Ref. \cite{Adamczewski-Musch:2020slf}, where the measurement of cumulants and factorial cumulants by the HADES Collaboration was reported. Another important effect is the global (or local) baryon number conservation, see, e.g.,
\cite{Bzdak:2012an,Braun-Munzinger:2016yjz,Rogly:2018kus,Braun-Munzinger:2019yxj,Acharya:2019izy,Vovchenko:2020tsr}. In Ref. \cite{Acharya:2019izy} the ALICE Collaboration emphasized the importance of the global baryon conservation at the LHC energies. 

In this paper we calculate the proton, antiproton, and the mixed proton-antiproton factorial cumulants up to the sixth order, assuming that the only source of correlations is the global conservation of baryon number. The factorial cumulants of the joint proton and antiproton multiplicity distribution $P(n_p, \bar{n}_p)$ contain more information than the cumulants of the net-proton distribution $P(n_p - \bar{n}_p)$ \cite{Bzdak:2016sxg}. Our results extend the so far published results and will allow for more sophisticated tests of the global baryon conservation effects in experiments.

In the next Section, we discuss our derivation of the proton, antiproton, and mixed proton-antiproton factorial cumulants. In Section III we present the exact results up to the sixth order and discuss some relations among them. We also provide very simple approximate expressions applicable at high energies. This is followed by the numerical results in Section IV. We finish the paper with comments and a summary. In Appendixes A-D some additional formulae and derivations are given.


\section{Calculation}
In this Section we derive analytically the factorial cumulants of proton and antiproton multiplicity distribution, originating from the global conservation of baryon number. We assume that the only source of correlations is given by the global conservation law. By $B$ we denote the conserved baryon number, $N_b$ and $\bar{N}_b$ are the event-by-event total numbers of baryons and anti-baryons, respectively, and $n_p$ and $\bar{n}_p$ are the numbers of observed protons and antiprotons in a given rapidity and/or transverse momentum interval.\footnote{Experimentally, one is usually restricted to the measurement of protons, however, the connection with baryons can be made \cite{Kitazawa:2011wh,Kitazawa:2012at}.}

The probability distribution of $n_p$ and $\bar{n}_p$ is given by\footnote{This derivation is slightly different than the one from Ref. \cite{Bzdak:2012an}, where the total volume was divided into observed and unobserved systems and the joint multiplicity distribution was written as a product of distributions from the two subvolumes (Eq. (5) in \cite{Bzdak:2012an}), see also \cite{Vovchenko:2020tsr}. Both procedures lead to identical results if the underlying distributions are Poissons.}
\begin{equation}
\begin{aligned} \label{eq:distr2}
P(n_{p}, \bar{n}_{p})&=A\sum_{N_{b}=n_{p}}^{\infty}\sum_{\bar{N}_b=\bar{n}_{p}}^{\infty}\delta_{N_{b}-\bar{N}_b,B}\left[\frac{\langle N_{b}\rangle^{N_{b}}}{N_{b}!}e^{-\langle N_{b}\rangle}\right] \left[\frac{\langle \bar{N}_b\rangle^{ \bar{N}_b }}{\bar{N}_b!}e^{-\langle \bar{N}_b \rangle}\right] \\ &\times \left[\frac{N_{b}!}{n_{p}!(N_{b}-n_{p})!}p^{n_{p}}(1-p)^{N_{b}-n_{p}}\right]\left[\frac{\bar{N}_b!}{\bar{n}_{p}!(\bar{N}_b-\bar{n}_{p})!}\bar{p}^{\bar{n}_{p}}(1-\bar{p})^{\bar{N}_b-\bar{n}_{p}}\right] ,
\end{aligned}
\end{equation}
where $p=\langle n_p \rangle / \langle N_b \rangle$ is the probability that the initial baryon is observed as a proton and $\bar{p}=\langle \bar{n}_p \rangle / \langle \bar{N}_b \rangle$ is the probability that the initial antibaryon is observed as an antiproton in a given acceptance region. $\langle x\rangle$ denotes an event average value of $x$.
The normalization constant is:
\begin{equation} \label{eq:normalization}
A = \frac{\left(\frac{\langle \bar{N}_b \rangle}{\langle N_{b} \rangle}\right)^{\frac{B}{2}} e^{^{\left\langle N_{b} \right\rangle + \left\langle \bar{N}_b \right\rangle}}} {I_{B}\left(2\sqrt{\left\langle N_{b} \right\rangle \left\langle \bar{N}_b \right\rangle}\right)},
\end{equation}
where $I_{\nu}(x)$ is a modified Bessel function of the order $\nu$. As already emphasized, our goal is to calculate the factorial cumulants assuming that the only source of correlation is given by the conservation of baryon number. Consequently, we start with $N_b$ and $\bar{N}_b$ following Poisson distributions and the multiplicities of observed protons and antiprotons are governed by binomial distributions \cite{Bzdak:2019pkr}, which do not introduce any new correlations (see also footnote $3$). The global baryon conservation is obviously enforced by $\delta_{N_{b}-\bar{N}_b,B}$.
Without this term, $P(n_{p}, \bar{n}_{p})$, would be given by a product of two Poisson distributions, and all the factorial cumulants would vanish. Note that Eq. (\ref{eq:distr2}) can be derived from a more general expression including protons, antiprotons, neutrons, and antineutrons. This is demonstrated in Appendix \ref{appendix:appA}.

Using Eqs. (\ref{eq:distr2}) and (\ref{eq:normalization}), it is straightforward to calculate the factorial moment generating function (a.k.a. probability generating function)
\begin{equation} \label{eq:hfun}
H(x, \bar{x})=\sum_{n_p=0}^{\infty}\sum_{\bar{n}_p=0}^{\infty}x^{n_p}\bar{x}^{\bar{n}_p} P(n_p, \bar{n}_p),
\end{equation} 
and the factorial cumulant generating function
\begin{equation}\label{eq:fcgf}
G(x, \bar{x})=\ln[H(x,\bar{x})].
\end{equation}
The result is:
\begin{equation}\label{eq:fcgf1}
G(x,\bar{x})=\ln\left[\left(\frac{px+1-p}{\bar{p}\bar{x}+1-\bar{p}}\right)^{\frac{B}{2}}\; \frac{I_{B}\left(2\sqrt{\left\langle N_{b} \right\rangle \left\langle \bar{N}_b \right\rangle(px+1-p)(\bar{p}\bar{x}+1-\bar{p})}\right)}{I_{B}\left(2\sqrt{\left\langle N_{b} \right\rangle \left\langle \bar{N}_b \right\rangle}\right)}\right].
\end{equation}

The factorial cumulants $\hat{C}^{(n,m)}$ which are the integrated (over a given acceptance region) correlation functions for (in our context) $n$ protons and $m$ antiprotons are given by
\begin{equation}\label{eq:fact-cum}
\hat{C}^{(n,m)}=\left. \frac{\partial^{n}}{\partial x^{n}}\frac{\partial^{m}}{\partial\bar{x}^{m}}G(x,\bar{x}) \right\rvert_{x=\bar{x}=1}.
\end{equation}

By definition, the factorial cumulants $\hat{C}^{(n,m)}=0$ for all $n \ge 1$, $m \ge 1$, if there are no correlations in the system \cite{Bzdak:2019pkr}, i.e., if $P(n_{p}, \bar{n}_{p})$ factorizes and both $n_{p}$ and $\bar{n}_{p}$ are distributed according to Poisson distributions. The global baryon number conservation, being a long-range correlation, results in non-zero $\hat{C}^{(n,m)}$. We note that the cumulants, which are usually measured in experiments, see, e.g., \cite{Adamczyk:2013dal,Luo:2015ewa,Adare:2015aqk,Behera:2018wqk,Adamczyk:2013dal,Acharya:2019izy,Adamczewski-Musch:2020slf,Adam:2020unf}, can be expressed by $\hat{C}^{(n,m)}$. We will discuss this issue later on. Here we only emphasize that the cumulants mix the factorial cumulants of different orders and in general, the factorial cumulants contain more information than the cumulants. 

Before we present our results let us introduce additional notation: 
\begin{equation} \label{eq:z-eq}
z=\sqrt{\langle N_{b}\rangle \langle \bar{N}_b\rangle},
\end{equation}
\begin{equation} \label{eq:Nbc}
\langle N_b \rangle _c = z \frac{I_{B-1}(2z)}{I_B(2z)}, \quad 
\langle \bar{N}_b \rangle _c = z \frac{I_{B+1}(2z)}{I_B(2z)} ,
\end{equation}
\begin{equation} \label{eq:zc}
z_{c} = \sqrt{\langle N_{b}\rangle _c \langle \bar{N}_b\rangle_c},
\end{equation}
where $\langle N_b \rangle$ is the mean number of baryons (present in Eq. (\ref{eq:distr2})) before the baryon number conservation is enforced, and $\langle N_b \rangle _c$ is the mean number of baryons with the conservation of baryon number (and analogously for antibaryons). The baryon number conserved averages obviously satisfy $\langle N_b \rangle _c - \langle \bar{N}_b\rangle_c = B$ (see Eq. (\ref{eq:Nbc}) and footnote 4).

\section{Results}
\subsection{Exact formulae}

In this Section we present analytic expressions for $\hat{C}^{(n,m)}$ up to the sixth order. It is natural to define:
\begin{equation}
\langle N\rangle_{c} = \langle N_{b}\rangle _c+\langle \bar{N}_b\rangle_c \,,
\end{equation}
which is the total average number of baryons. To present the formulae in a more compact way we identified commonly appearing terms and denoted them as:
\begin{equation} \label{eq:delta}
\Delta=z_{c}^{2}-z^{2}\,,
\end{equation}
\begin{equation} \label{eq:gamma}
\gamma=z_{c}^{2}+\Delta \langle N\rangle_{c} \,,
\end{equation}
\begin{equation} \label{eq:gamma}
\be=\gamma(\langle N\rangle_{c} + 2) + 2\Delta^2 \,,
\end{equation}
where $\langle N\rangle_{c}$, $\Delta$, $\gamma$ and $\be$ depend on $B$ and $z$ only, 
see Eqs. (\ref{eq:Nbc}) and (\ref{eq:zc}). 
The factorial cumulants read\footnote{In this calculation we extensively use $I_{\nu-1}(x) - I_{\nu+1}(x) = \frac{2\nu}{x} I_{\nu}(x)$.}:
{ 
\setlength{\abovedisplayskip}{6pt}
\setlength{\belowdisplayskip}{6pt}
\begin{fleqn}
	\begin{gather}
 	\begin{aligned} \label{eq:c10}
 		\hat{C}^{(1,0)}&=p\langle N_{b}\rangle _c 
 	\end{aligned}
 	\end{gather}
	\begin{gather}
 	\begin{aligned} \label{eq:c20}
 		\hat{C}^{(2,0)}&=-p^{2}\left(\langle N_{b}\rangle _c+\Delta\right)
 	\end{aligned} \\
 	\begin{aligned}
 		\hat{C}^{(1,1)}&=-p\bar{p}\Delta
 	\end{aligned}
 	\end{gather}
	\begin{gather}
 	\begin{aligned} \label{eq:c30}
 		\hat{C}^{(3,0)}&=p^{3}\left[2!\left(\langle N_{b}\rangle_{c}+\Delta + \tfrac{1}{2}\gamma\right)\right]
 	\end{aligned} \\
 	\begin{aligned} \label{eq:c21}
 		\hat{C}^{(2,1)}&=p^{2}\bar{p}\, \gamma
 	\end{aligned}
 	\end{gather}
	\begin{gather}
 	\begin{aligned} \label{eq:c40}
 		\hat{C}^{(4,0)}&=-p^{4}\left[3!\left(\langle N_{b}\rangle_{c}+\Delta+\tfrac{1}{2}\gamma\right) + \be\right]
 	\end{aligned} \\
 	\begin{aligned} \label{eq:c31}
 		\hat{C}^{(3,1)}&=-p^{3}\bar{p}\be
 	\end{aligned} \\
 	\begin{aligned} \label{eq:c22}
 		\hat{C}^{(2,2)}&=-p^{2}\bar{p}^{2}\left(\be - \gamma \right)
 	\end{aligned}
 	\end{gather}
 	\begin{gather}
 	\begin{aligned} \label{eq:c50}
 		\hat{C}^{(5,0)}&=p^{5}\left[4!\left(\langle N_{b}\rangle_{c}+\Delta + \tfrac{1}{2}\gamma\right)+(\langle N\rangle_{c}+7)\be + 6\gamma\Delta\right]
 	\end{aligned} \\
 	\begin{aligned} \label{eq:c41}
 		\hat{C}^{(4,1)}&=p^{4}\bar{p}\left[(\langle N\rangle_{c}+3)\be + 6\gamma\Delta\right] 
 	\end{aligned} \\
 	\begin{aligned} \label{eq:c32}
 		\hat{C}^{(3,2)}&=p^{3}\bar{p}^{2}\left[(\langle N\rangle_{c}+1)\be + 6\gamma\Delta\right] 
 	\end{aligned}
 	\end{gather}
	\begin{gather}
 	\begin{aligned} \label{eq:c60}
 		\hat{C}^{(6,0)} &=-p^{6}\left[ 5!\left( \langle N_{b}\rangle _{c}+\Delta +
        \tfrac{1}{2}\gamma \right) +\left\{ (\langle N\rangle _{c}+5)(\langle
        N\rangle _{c}+7)+12\right\} \beta +6\gamma ^{2}+16\Delta ^{3} \right. \\
        &\left. +2\gamma \Delta (7\langle N\rangle_{c}+35)\right] 
 	\end{aligned} \\
 	\begin{aligned} \label{eq:c51}
 		\hat{C}^{(5,1)}&=-p^{5}\bar{p}\left[(\langle N\rangle_{c}+3)(\langle N\rangle_{c}+4)\be+6\gamma^{2}+16\Delta^{3}+2\gamma\Delta(7\langle N\rangle_{c}+20)\right] 
 	\end{aligned} \\
 	\begin{aligned} \label{eq:c42}
 		\hat{C}^{(4,2)}&=-p^{4}\bar{p}^{2}\left[(\langle N\rangle_{c}+1)(\langle N\rangle_{c}+3)\be+6\gamma^{2}+16\Delta^{3}+2\gamma\Delta(7\langle N\rangle_{c}+11)\right] 
 	\end{aligned} \\
 	\begin{aligned} \label{eq:c33}
 		\hat{C}^{(3,3)}&=-p^{3}\bar{p}^{3}\left[(\langle N\rangle_{c}+1)(\langle N\rangle_{c}+2)\be+6\gamma^{2}+16\Delta^{3}+2\gamma\Delta(7\langle N\rangle_{c}+8)\right]
 	\end{aligned}
	\end{gather}
\end{fleqn}
}

Having $\hat{C}^{(n,m)}$, one can easily obtain $\hat{C}^{(m,n)}$: 
\begin{equation}
\hat{C}^{(m,n)} = \hat{C}^{(n,m)}\left(p\rightarrow\bar{p}, \bar{p}\rightarrow p\right) 
\quad \text{for} \quad n\,m \ne 0 \,,
\end{equation}
\begin{equation}
\hat{C}^{(0,n)} = \hat{C}^{(n,0)}\left(p\rightarrow\bar{p}, \langle N_b\rangle_c \rightarrow \langle \bar{N}_b\rangle_c \right) \,,
\end{equation}
that is, to obtain $\hat{C}^{(m,n)}$ from $\hat{C}^{(n,m)}$ with both $n$ and $m$ larger than zero, it is enough to exchange $p$ with $\bar{p}$. To obtain $\hat{C}^{(0,n)}$ from $\hat{C}^{(n,0)}$ it is also necessary to replace $\langle N_{b} \rangle _c$ by $\langle \bar{N}_b\rangle_c$. 
For example, $\hat{C}^{(0,1)}=\bar{p}\langle \bar{N}_b\rangle_c$ and $\hat{C}^{(1,2)}=p\bar{p}^2\gamma$.

\subsection{Relations}
As seen from \Crefrange{eq:c10}{eq:c33}, $\hat{C}^{(n,m)}$ is proportional to $p^{n}\bar{p}^{m}$.\footnote{This is not unexpected. As argued in, e.g., Refs. \cite{Bzdak:2016sxg,Bzdak:2016jxo} the long-range correlation, such as global baryon conservation, naturally results in $\hat{C}^{(n,m)}$ being proportional to $\langle n_p \rangle^n \langle \bar{n}_{p} \rangle^m$, 
where $\langle n_p \rangle = p \langle N_b \rangle$ and 
$\langle \bar{n}_p \rangle = \bar{p} \langle \bar{N}_b \rangle$.}
Therefore it is natural to study the following ratios
\begin{equation} \label{eq:rhat}
\hat{R}^{(n,m)}=\frac{\hat{C}^{(n,m)}}{p^{n}\bar{p}^{m}}\,,
\end{equation}
which are independent of the size of the chosen acceptance bin.

Using \Crefrange{eq:c10}{eq:c33} we find several simple relations between various $\hat{R}^{(n,m)}$:
\begin{fleqn}
\begin{gather}
\begin{aligned}
	\hat{R}^{(2,0)}=\hat{R}^{(1,1)} - \hat{R}^{(1,0)}\,,
\end{aligned} \\
\begin{aligned}
	\hat{R}^{(3,0)}=\hat{R}^{(2,1)} - 2\hat{R}^{(2,0)}\,,
\end{aligned} \\
\begin{aligned}
	\hat{R}^{(4,0)}=\hat{R}^{(3,1)} - 3\hat{R}^{(3,0)}\,,
\end{aligned} \\
\begin{aligned}
	\hat{R}^{(5,0)}=\hat{R}^{(4,1)} - 4\hat{R}^{(4,0)}\,,
\end{aligned} \\
\begin{aligned}
	\hat{R}^{(6,0)}=\hat{R}^{(5,1)} - 5\hat{R}^{(5,0)}\,,
\end{aligned} \\
\begin{aligned}
	\hat{R}^{(3,1)}=\hat{R}^{(2,2)} - \hat{R}^{(2,1)}\,,
\end{aligned} \\
\begin{aligned}
	\hat{R}^{(4,1)}=\hat{R}^{(3,2)} - 2\hat{R}^{(3,1)}\,,
\end{aligned} \\
\begin{aligned}
	\hat{R}^{(5,1)}=\hat{R}^{(4,2)} - 3\hat{R}^{(4,1)}\,,
\end{aligned} \\
\begin{aligned}
	\hat{R}^{(4,2)}=\hat{R}^{(3,3)} - \hat{R}^{(3,2)}\,,
\end{aligned}
\end{gather}
\end{fleqn}
or in general ($n>0$ or $m>0$)
\begin{equation}
   \hat{R}^{(n+1,m)} = \hat{R}^{(n,m+1)} - (n-m)\hat{R}^{(n,m)} \,,
\end{equation}
which we verified by direct calculations up to $n+m<9$.

\subsection{Approximate formulae for $B=0$}
\label{sec:Beq0}
Here, we analyze in detail the special case of $B=0$, meaning the same total number of baryons and antibaryons, which characterizes large energy conditions, such as at the LHC CERN. In this case $\langle N_b\rangle_c = \langle \bar{N}_b\rangle_c$, 
$z_c = \langle N_b\rangle_c$ and $\langle N\rangle_c = 2\langle N_b\rangle_c$. 
All components appearing in \Crefrange{eq:c10}{eq:c33}, that is, $\langle N\rangle_c$, $\langle N_b\rangle_c$, $\Delta$, $\gamma$ and $\be$ depend on $z$ only. 
Next, we apply to Eq. (\ref{eq:Nbc}) the asymptotic (large argument) expansion of the modified Bessel function \cite{abramowitz-stegun}:
\begin{equation} \label{eq:asympt-bessel}
I_{\nu}(x)\sim \frac{e^{x}}{\sqrt{2\pi x}}\left(1+\sum_{n=1}^{\infty}\frac{(-1)^{n}\prod_{i=1}^{n}(4\nu^{2}-(2i-1)^2)}{n!(8x)^n}\right).
\end{equation}

After eliminating the Bessel functions (the higher the order of the factorial cumulant, the more terms are needed in Eq. (\ref{eq:asympt-bessel})) we expand $\hat{R}^{(n,m)}(z)$ into a power series\footnote{Here we introduce $z=1/y$ and expand about $y=0$ and then substitute back $y=1/z$.} for large $z$ and obtain the dependency of the form
\begin{equation}
\hat{R}^{(n,m)}(z) \sim a_1 z + a_0 + a_{-1} z^{-1} + a_{-2} z^{-2} + ... \,,
\end{equation}
where the coefficients $a_i$ depend on $n$ and $m$. It is worth noting that $\hat{R}^{(n,m)}(z)$ grows linearly with $z$ for large $z$. The details and explicit expressions for $\hat{R}^{(n,m)}(z)$ are presented in Appendix \ref{appendix:appE}.

It can be proved (see Appendix \ref{appendix:appE}) that $\hat{R}^{(n,m)}(z_c)$ is also of the same form, that is, the highest-order term is proportional to $z_c$ and the coefficients of the series can be easily calculated. The obtained asymptotic expressions for $\hat{R}^{(n,m)}(z_c)$ at large $z_c$ are given below ($z_{c}=\langle N_b \rangle _c = \langle \bar{N}_b \rangle _c$):

{ 
\setlength{\abovedisplayskip}{6pt}
\setlength{\belowdisplayskip}{6pt}
\begin{fleqn}
\begin{gather}
 	\begin{aligned} \label{eq:r20-of-zc}
 		\hat{R}^{(2,0)}(z_c) \sim -\tfrac{1}{2}z_c + \tfrac{1}{8} + \tfrac{1}{32}z_c^{-1} + ...
 	\end{aligned} \\
 	\begin{aligned}
 		\hat{R}^{(1,1)}(z_c) \sim \tfrac{1}{2}z_c + \tfrac{1}{8}  + \tfrac{1}{32}z_c^{-1} + ...
 	\end{aligned}
\end{gather}
\begin{gather}
 	\begin{aligned}
 		\hat{R}^{(3,0)}(z_c) \sim \tfrac{3}{4}z_c - \tfrac{5}{16} - \tfrac{3}{32}z_c^{-1} + ...
 	\end{aligned} \\
 	\begin{aligned}
 		\hat{R}^{(2,1)}(z_c) \sim -\tfrac{1}{4}z_c - \tfrac{1}{16}  - \tfrac{1}{32}z_c^{-1} + ...
 	\end{aligned} 
\end{gather}
\begin{gather}
 	\begin{aligned}
 		\hat{R}^{(4,0)}(z_c) \sim -\tfrac{15}{8}z_c + \tfrac{33}{32} + \tfrac{45}{128}z_c^{-1} + ...
 	\end{aligned} \\
 	\begin{aligned}
 		\hat{R}^{(3,1)}(z_c) \sim \tfrac{3}{8}z_c + \tfrac{3}{32} + \tfrac{9}{128}z_c^{-1} + ...
 	\end{aligned} \\
 	\begin{aligned}
 		\hat{R}^{(2,2)}(z_c) \sim \tfrac{1}{8}z_c + \tfrac{1}{32} + \tfrac{5}{128}z_c^{-1} + ...
 	\end{aligned} 
\end{gather}
\begin{gather}
 	\begin{aligned}
 		\hat{R}^{(5,0)}(z_c) \sim \tfrac{105}{16}z_c  - \tfrac{279}{64} - \tfrac{105}{64}z_c^{-1} + ...
 	\end{aligned} \\
 	\begin{aligned}
 		\hat{R}^{(4,1)}(z_c) \sim -\tfrac{15}{16}z_c - \tfrac{15}{64} - \tfrac{15}{64}z_c^{-1} + ...
 	\end{aligned} \\
 	\begin{aligned}
 		\hat{R}^{(3,2)}(z_c) \sim -\tfrac{3}{16}z_c  - \tfrac{3}{64} - \tfrac{3}{32}z_c^{-1} + ...
 	\end{aligned} 
\end{gather}
\begin{gather}
 	\begin{aligned}
 		\hat{R}^{(6,0)}(z_c) \sim -\tfrac{945}{32}z_c + \tfrac{2895}{128} + \tfrac{4725}{512}z_c^{-1} + ...
 	\end{aligned} \\
 	\begin{aligned}
 		\hat{R}^{(5,1)}(z_c) \sim \tfrac{105}{32}z_c  + \tfrac{105}{128} + \tfrac{525}{512}z_c^{-1} + ...
 	\end{aligned} \\
 	\begin{aligned}
 		\hat{R}^{(4,2)}(z_c) \sim \tfrac{15}{32}z_c + \tfrac{15}{128} + \tfrac{165}{512}z_c^{-1} + ...
 	\end{aligned} \\
 	\begin{aligned}\label{eq:r33-of-zc}
 		\hat{R}^{(3,3)}(z_c) \sim \tfrac{9}{32}z_c + \tfrac{9}{128} + \tfrac{117}{512}z_c^{-1} + ...
 	\end{aligned} 
\end{gather}
\end{fleqn}
}

We checked, see Section IV, that the obtained approximate formulae work with very good accuracy already from $z_c = \langle N_b \rangle _c > 2$.

\section{Numerical results}
In this Section we present numerical results for 
$\hat{R}^{(n,m)}(z_c) = \hat{C}^{(n,m)}/(p^{n}\bar{p}^{m})$ 
for two special cases: $B=0$ corresponding to large energies, and $B=300$ corresponding to central collisions at low energies in heavy-ion collisions.

\subsection{$B=0$}
\begin{figure}[H]
\begin{center}
	\includegraphics[width=0.49\textwidth]{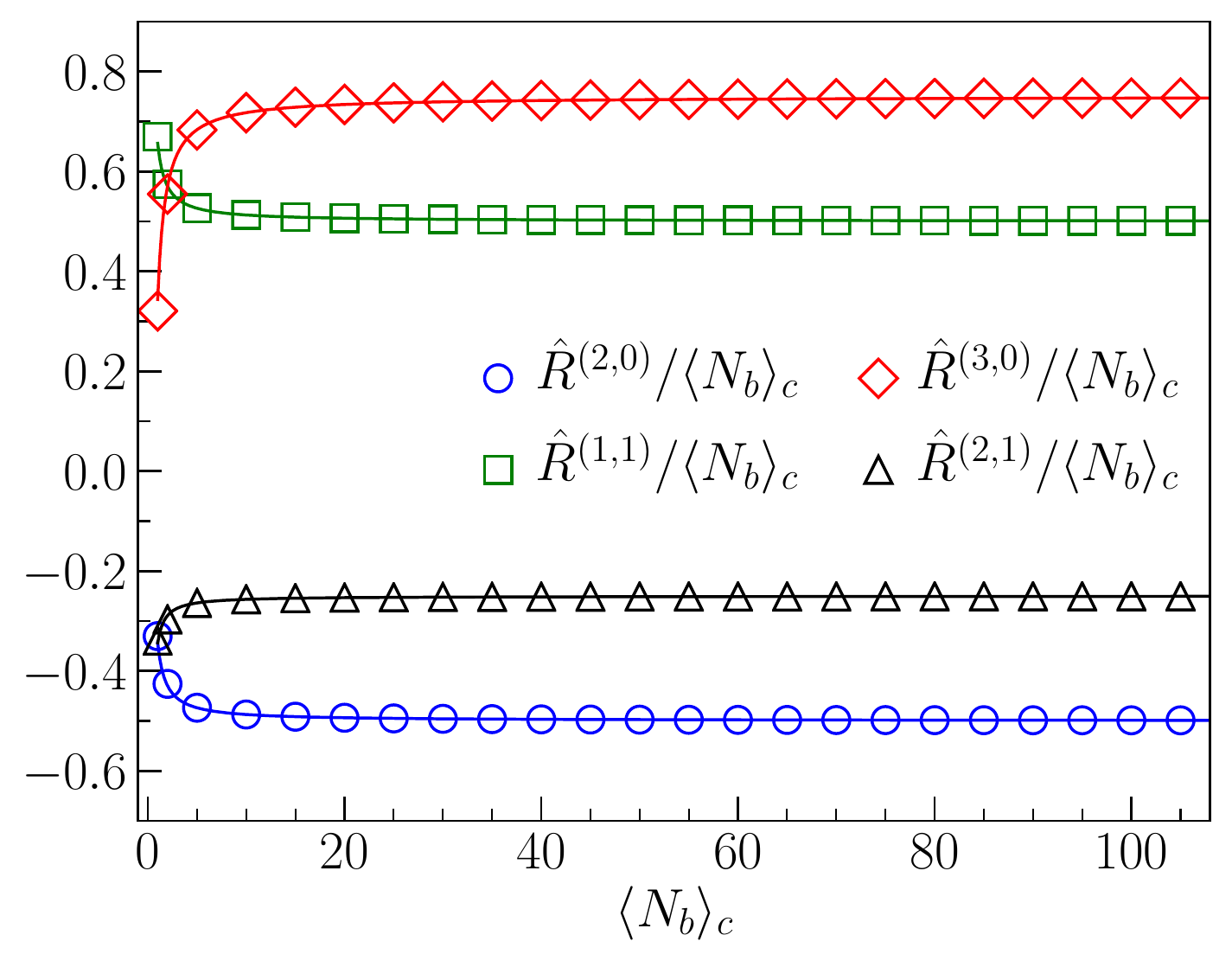}
	\includegraphics[width=0.49\textwidth]{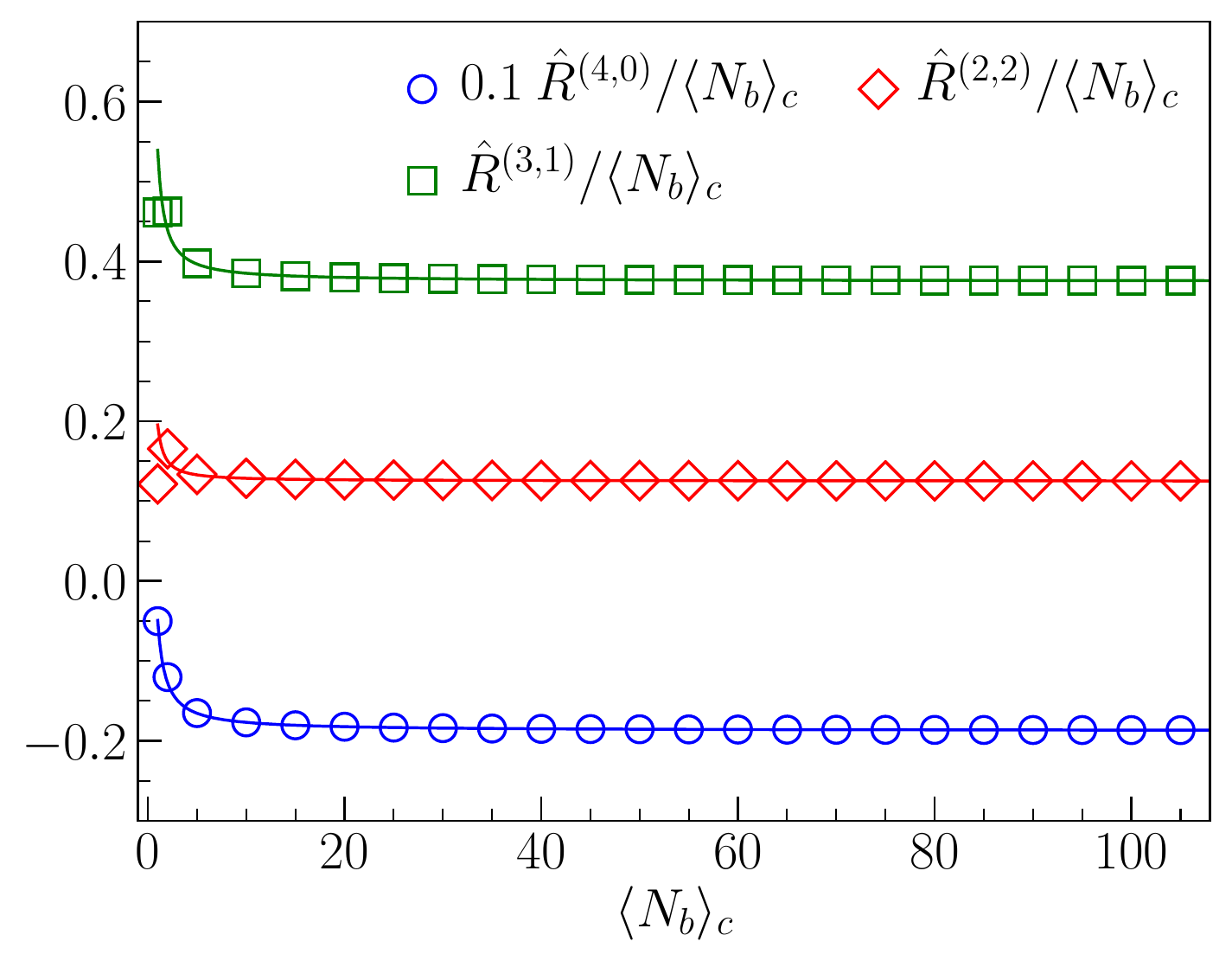}
	\includegraphics[width=0.49\textwidth]{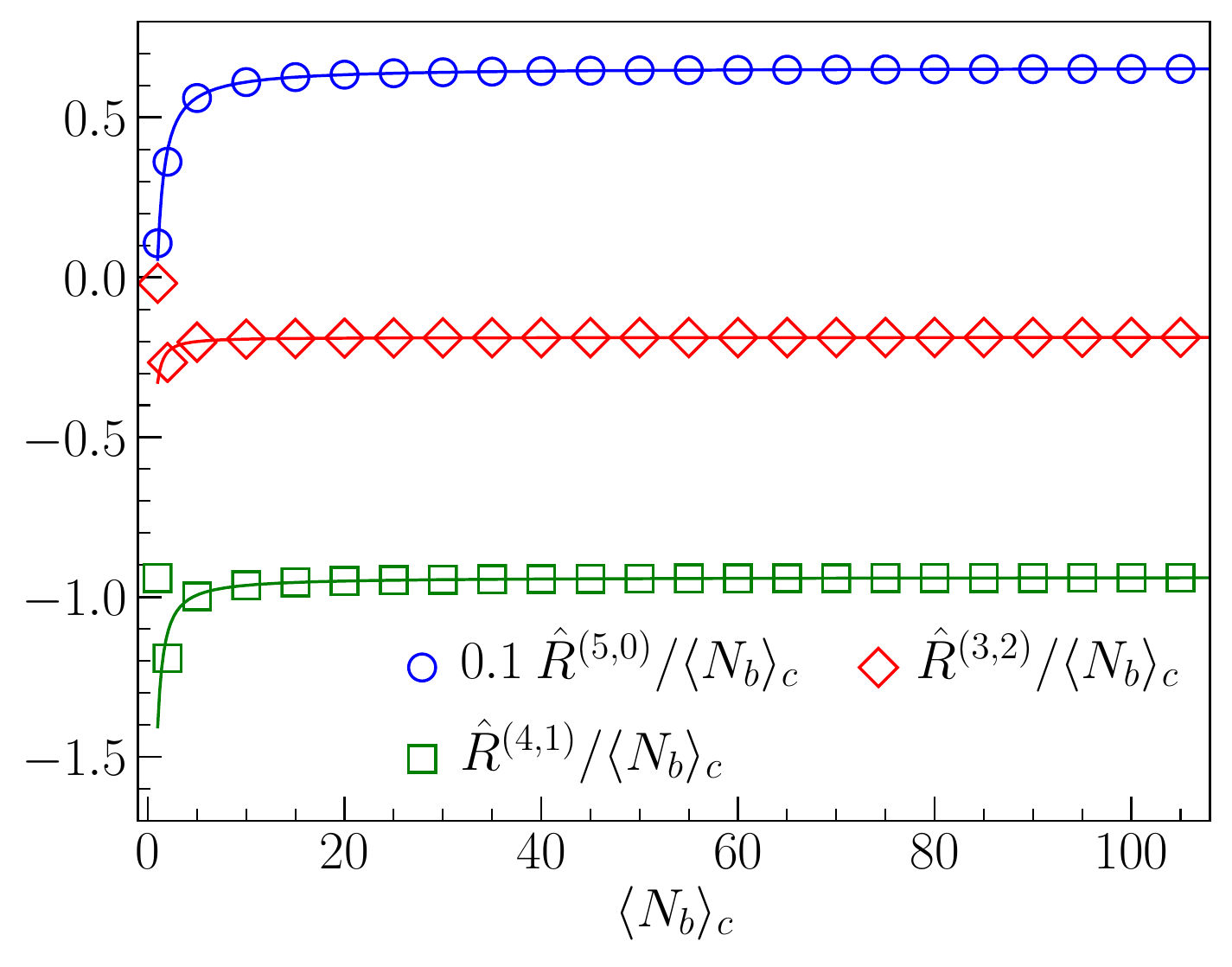}
	\includegraphics[width=0.49\textwidth]{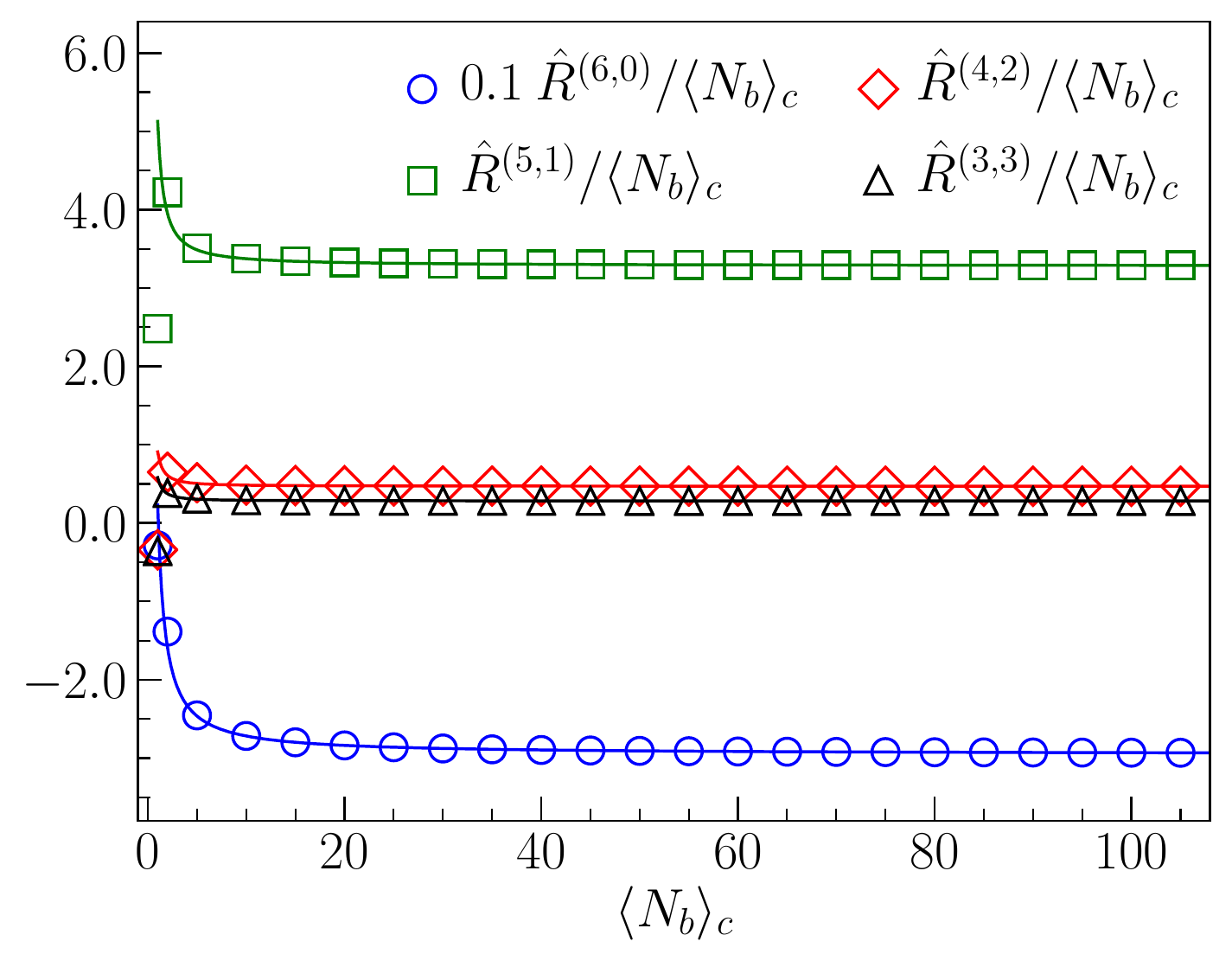}
\caption{$\hat{R}^{(n,m)}/\langle N_b \rangle _c$ as a function of $\langle N_b \rangle _c$ for $B=0$, where $\hat{R}^{(n,m)} = \hat{C}^{(n,m)}/(p^n\bar{p}^m)$.  
Markers represent exact formulae for the factorial cumulants $\hat{C}^{(n,m)}$ given by \Crefrange{eq:c10}{eq:c33}, whereas lines represent our asymptotic formulae (large $\langle N_b \rangle _c$) given by \Crefrange{eq:r20-of-zc}{eq:r33-of-zc}. 
Markers are plotted for $\langle N_b \rangle _c$ = 1, 2, 5, 10, 15, .... For $\langle N_b \rangle _c > 2$ the approximated formulae work very well, achieving precision better than 1\% starting from $\langle N_b \rangle _c $ between 2 and 7 depending on the order of the factorial cumulant. 
Some of the functions were scaled by a factor of 0.1 to improve readability.} \label{fig:r-100}
\end{center}
\end{figure}

For $B=0$, $z_c=\langle N_b \rangle _c = \langle \bar{N}_b \rangle _c = \langle N \rangle _c /2$ and therefore $\hat{R}^{(n,m)}(z_c)$ equals $\hat{R}^{(n,m)}(\langle N_b \rangle _c)$. From \Crefrange{eq:r20-of-zc}{eq:r33-of-zc} it is clear that the dominant contribution is linear with $z_c=\langle N_b \rangle _c$ and there are certain deviations for small $\langle N_b \rangle _c$. Therefore, for $B=0$, it is natural to divide $\hat{R}^{(n,m)}$ by $\langle N_b \rangle _c$ so that the leading term is simply constant.
In Fig. \ref{fig:r-100} we present $\hat{R}^{(n,m)}(\langle N_b \rangle _c)$ divided by $\langle N_b \rangle _c$ for all the discussed factorial cumulants.
Markers represent exact formulae for the factorial cumulants $\hat{C}^{(n,m)}$ given by \Crefrange{eq:c10}{eq:c33}, whereas lines represent our asymptotic expressions (large $\langle N_b \rangle _c$) given by \Crefrange{eq:r20-of-zc}{eq:r33-of-zc}.\footnote{For the exact results we first take $\langle N_b \rangle _c$ and solve Eq. (\ref{eq:Nbc}) for $z$, which we substitute to \Crefrange{eq:c10}{eq:c33}.}
These functions are essentially constant, in agreement with our asymptotic results, except for small values of $\langle N_b \rangle _c$.
The approximated formulae work very well starting from $\langle N_b \rangle _c \approx 2$. The precision better than 1\% is obtained starting from $\langle N_b \rangle _c \approx 7$ in the worst case of the sixth order factorial cumulants.

\subsection{$B=300$}
\begin{figure}[H]
\begin{center}
	\includegraphics[width=0.49\textwidth]{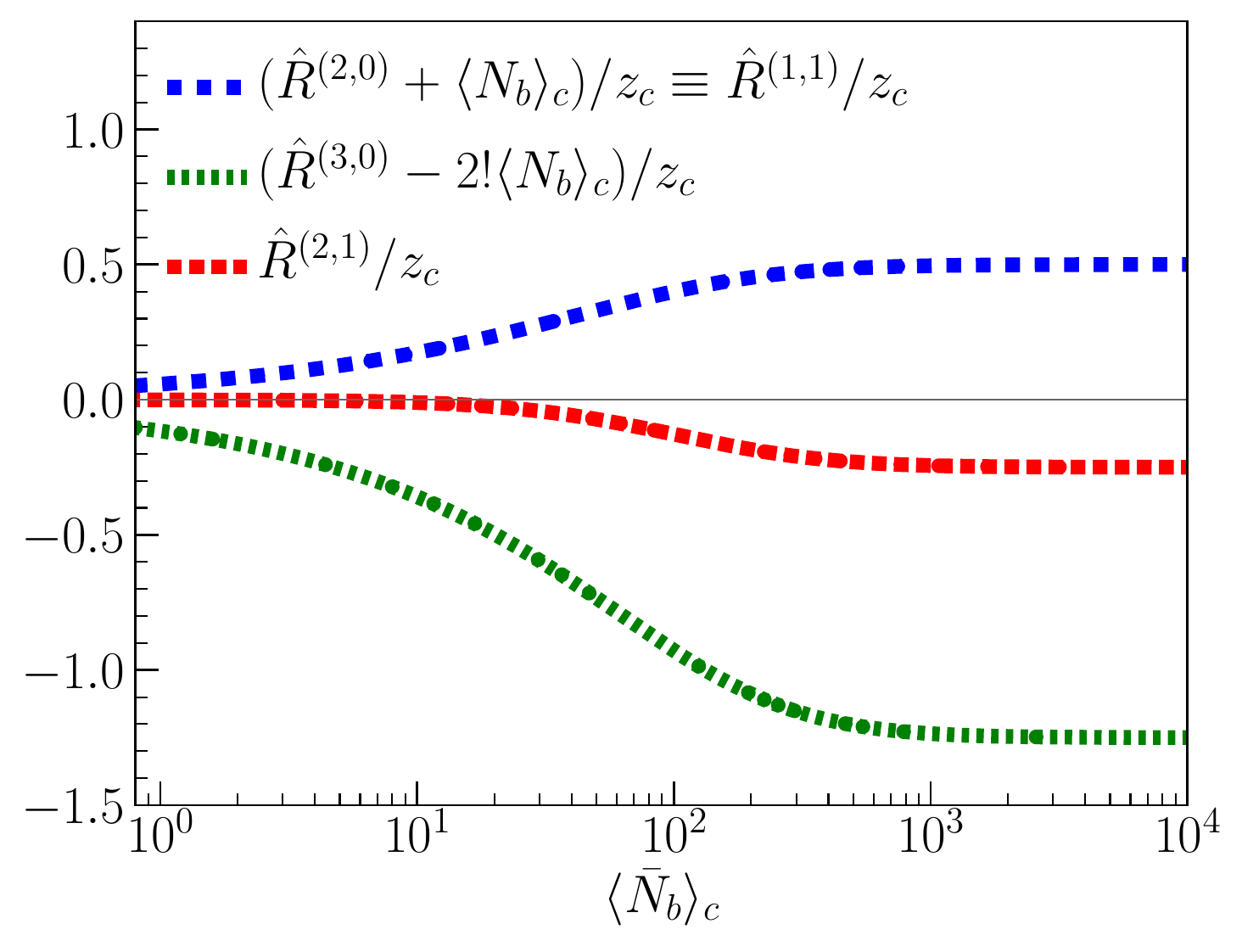}
	\includegraphics[width=0.49\textwidth]{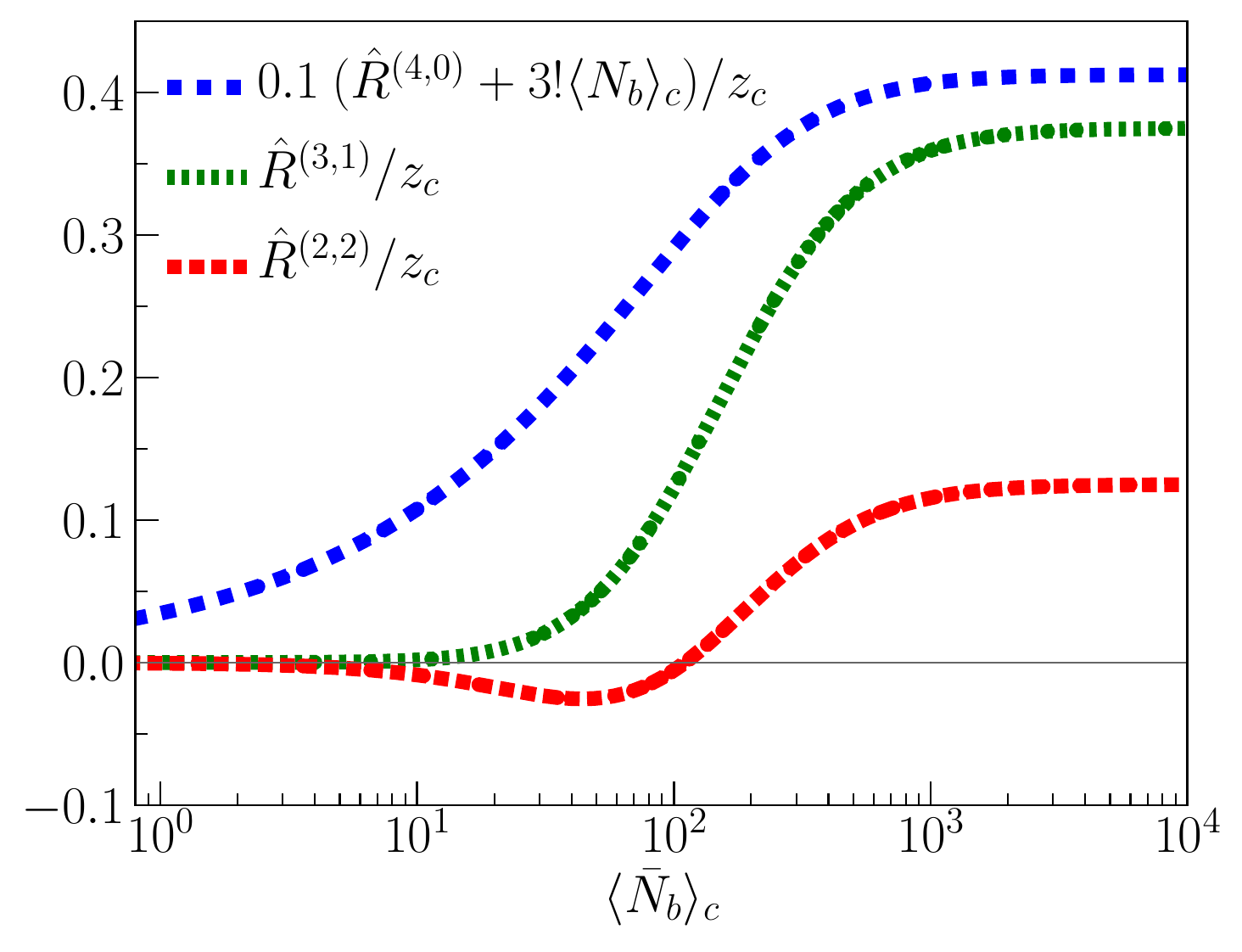}
	\includegraphics[width=0.49\textwidth]{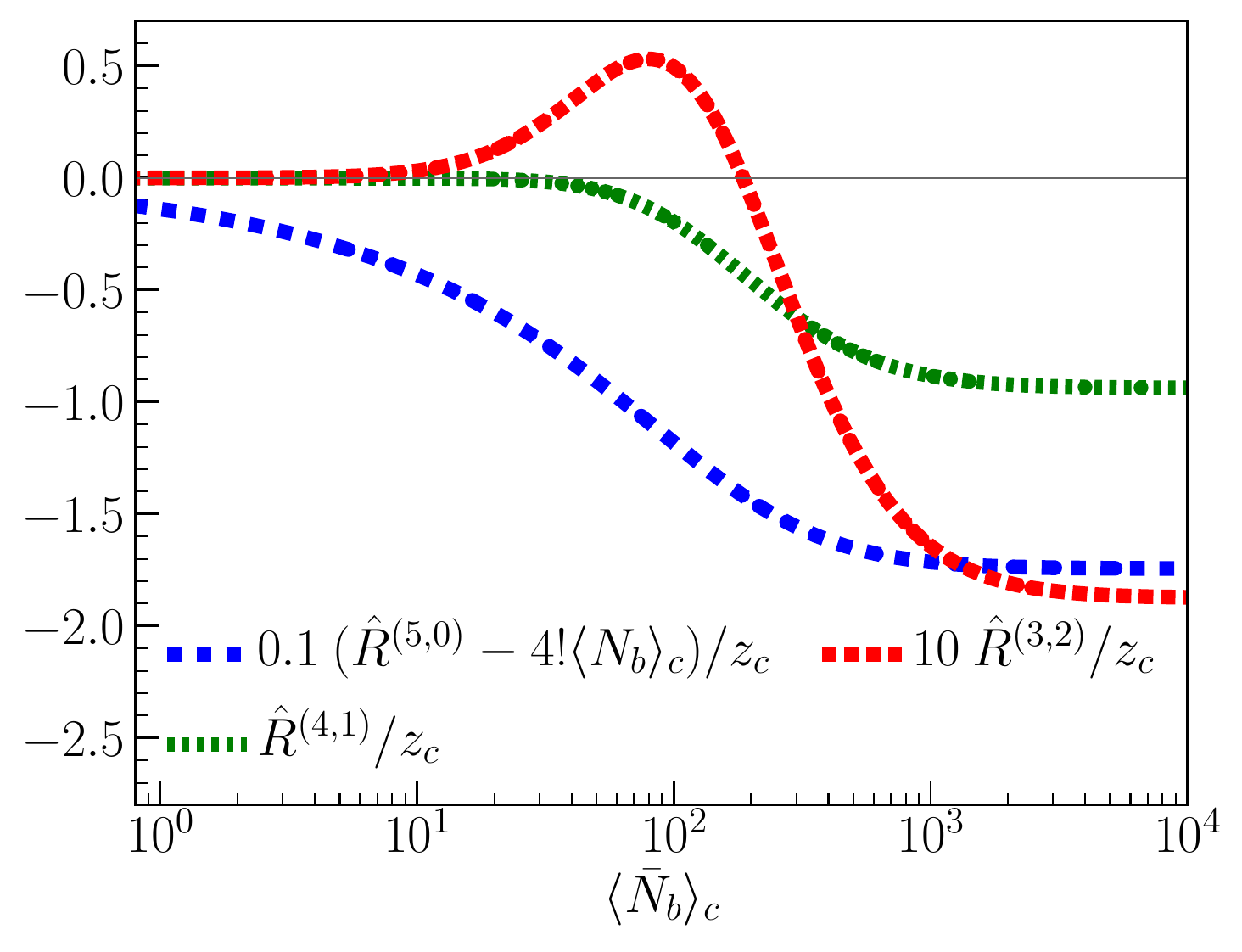}
	\includegraphics[width=0.49\textwidth]{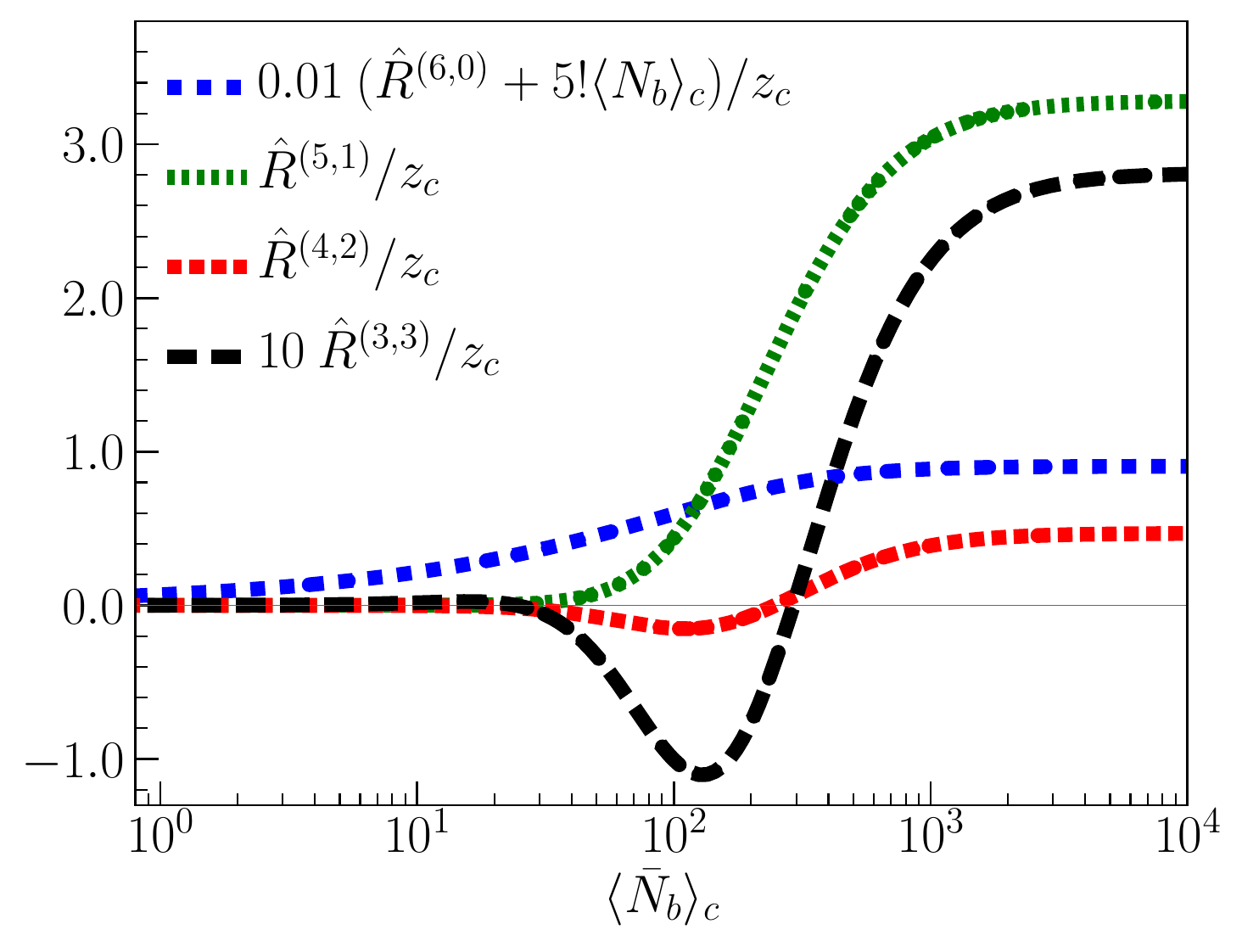}
\caption{$\hat{R}^{(n,m)}/z_c$ as a function of $\langle \bar{N}_b\rangle_c$ for $B=300$ based on \Crefrange{eq:c10}{eq:c33}. $\hat{R}^{(n,m)} = \hat{C}^{(n,m)}/(p^n\bar{p}^m)$. 
For $m=0$ we present $(\hat{R}^{(n,0)} - (-1)^{n-1}(n-1)! \langle N_b \rangle _c)/z_c$ because it gives the same values for both $\hat{R}^{(n,0)}$ and $\hat{R}^{(0,n)}$. 
Some of the functions were scaled by a factor of 10, 0.1 or 0.01 to improve readability.
Note the logarithmic scale on the horizontal axis.} \label{fig:r-300}
\end{center}
\end{figure}

\begin{figure}[H]
\begin{center}
	\includegraphics[width=0.49\textwidth]{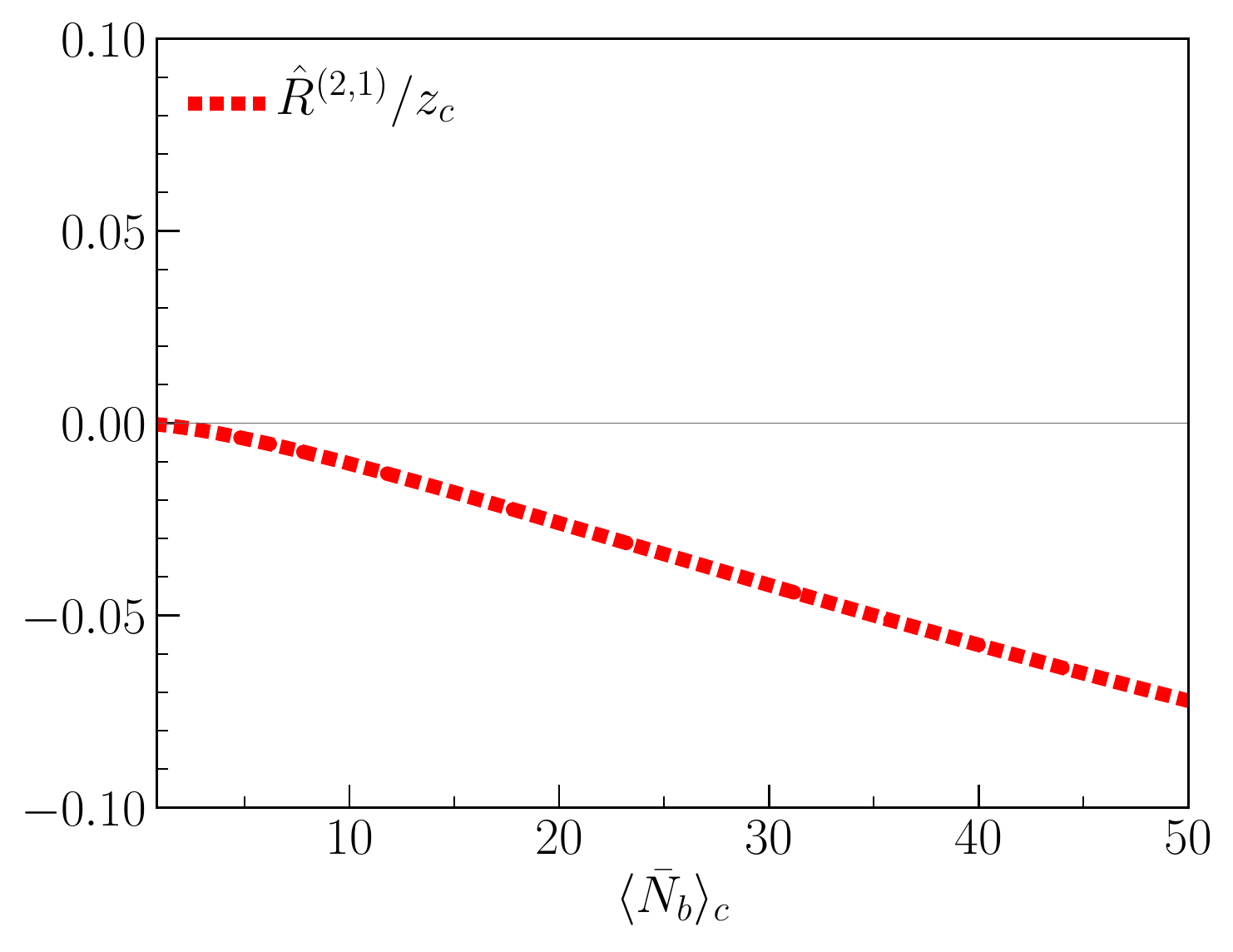}
	\includegraphics[width=0.49\textwidth]{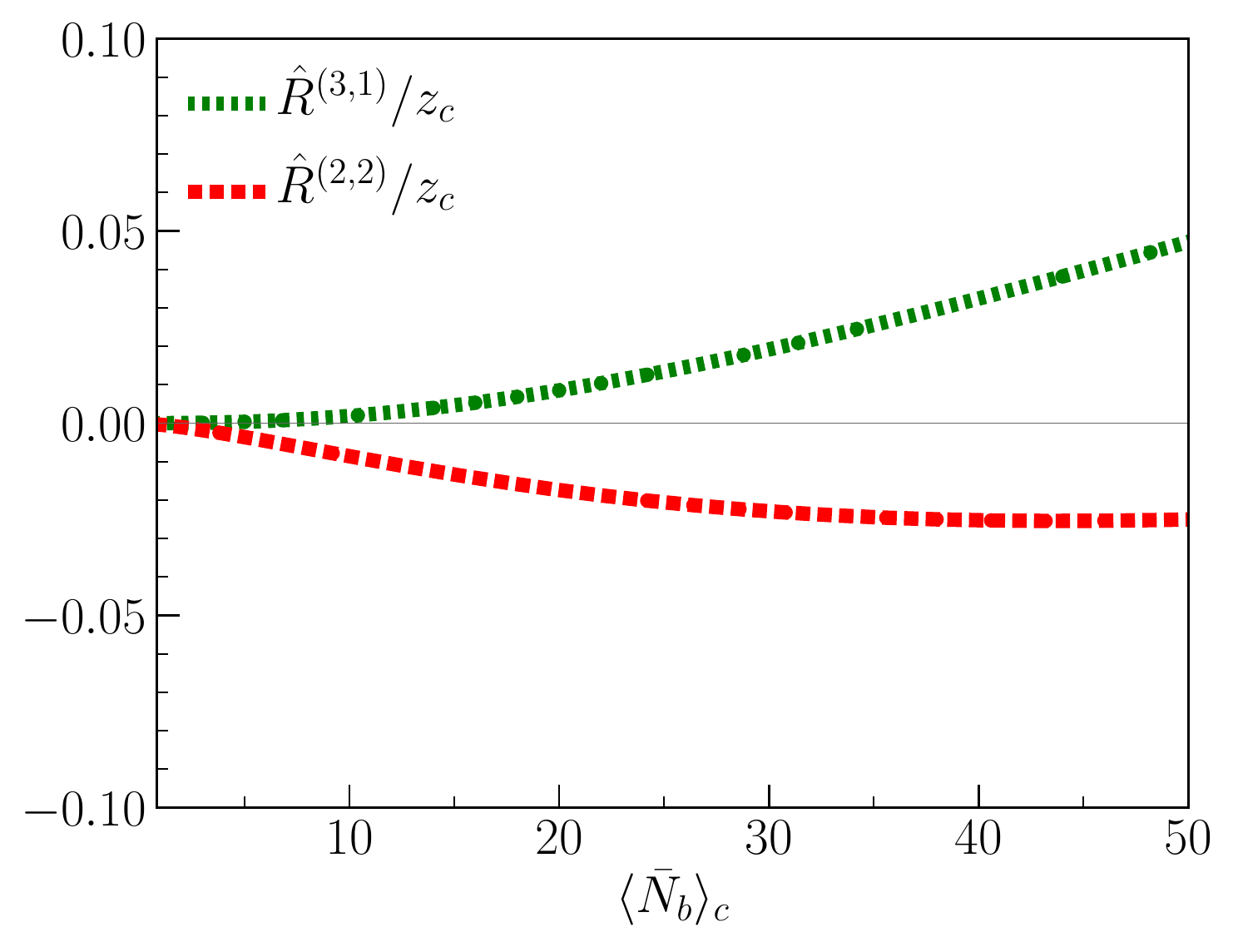}
	\includegraphics[width=0.49\textwidth]{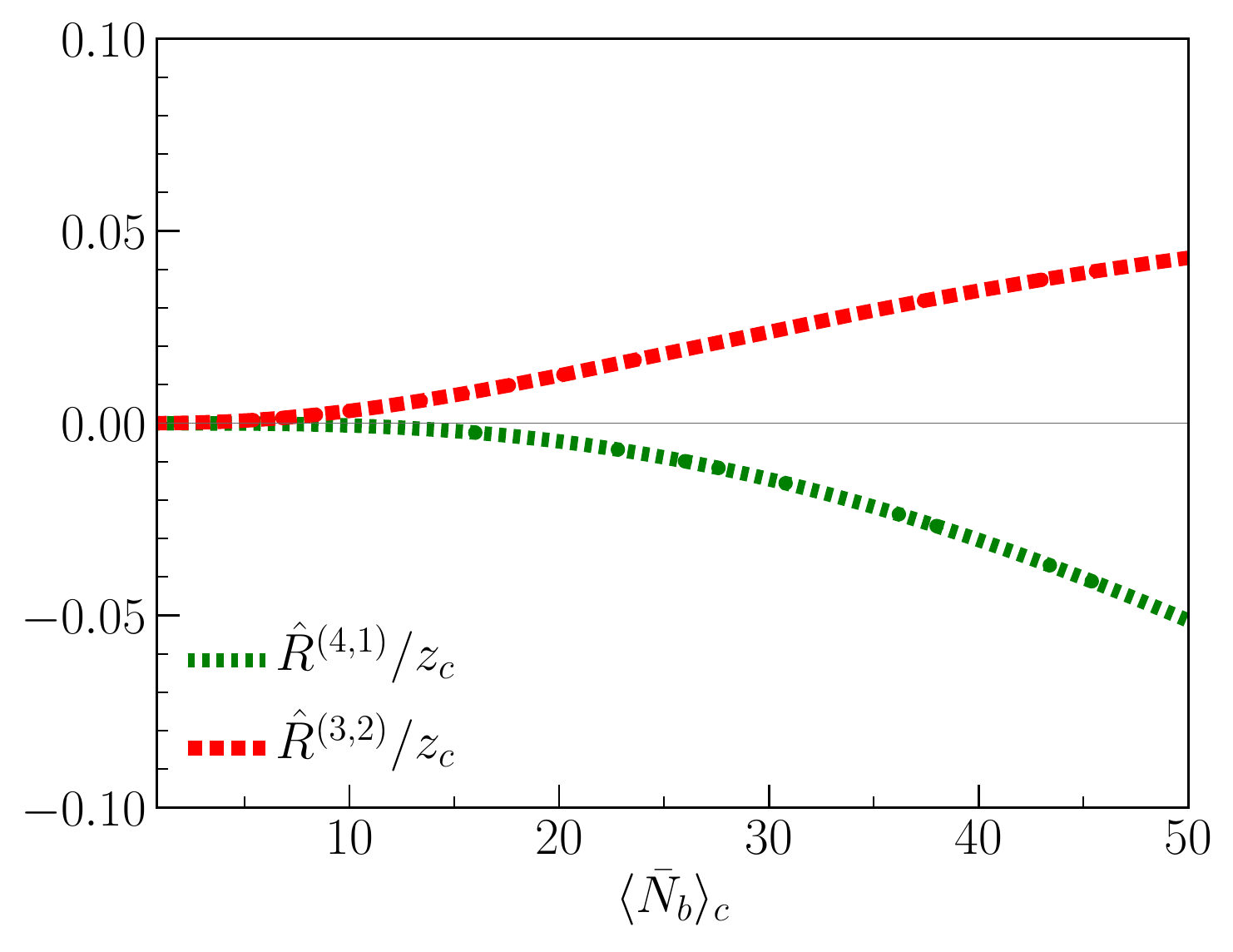}
	\includegraphics[width=0.49\textwidth]{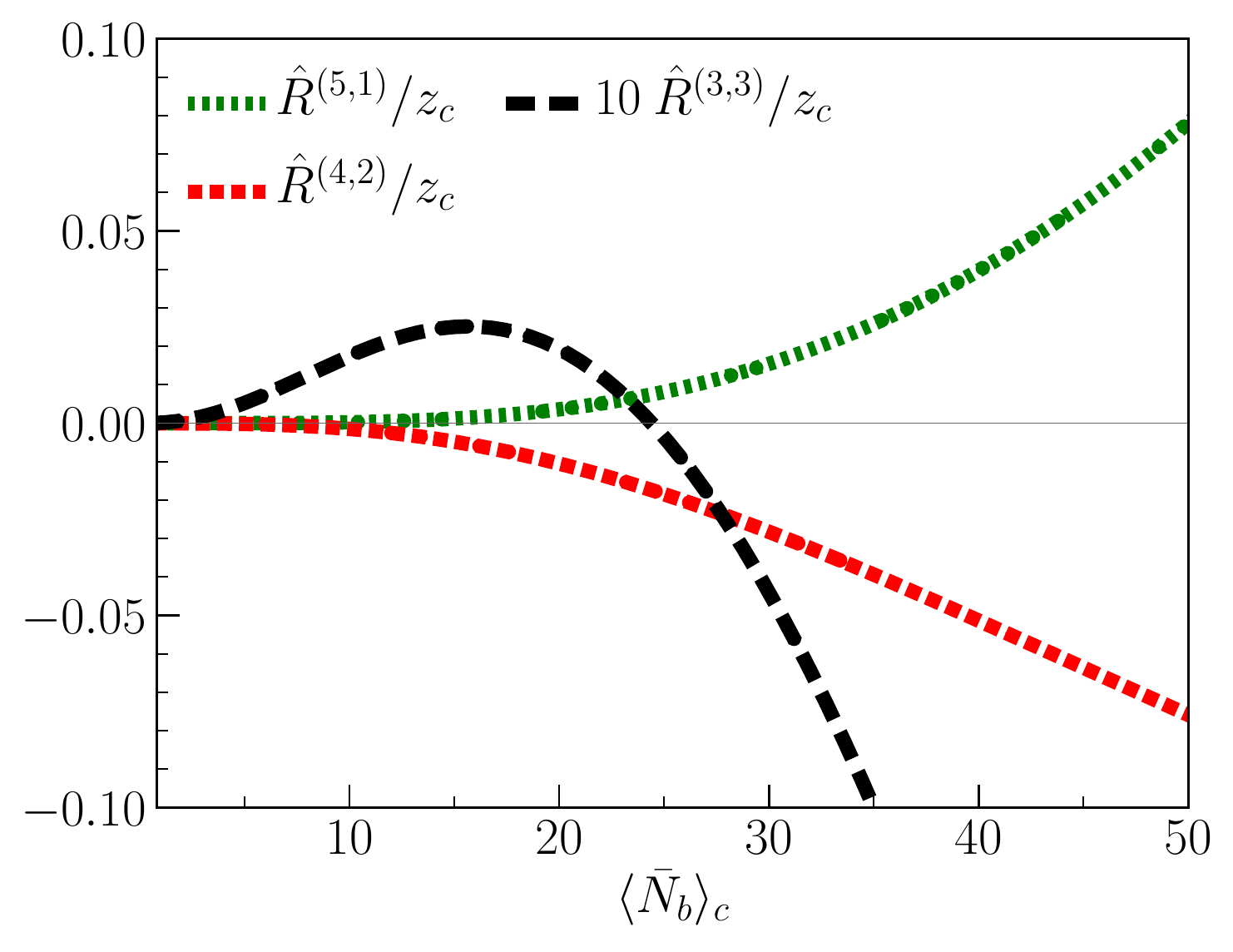}
\caption{Same as Fig. \ref{fig:r-300} but for $nm \neq 0$ and for small $\langle \bar{N}_b\rangle_c$. Note the linear scale on the horizontal axis. $\hat{R}^{(3,3)}/z_c$ was scaled by 10 to make the maximum at $\langle \bar{N}_b\rangle_c \approx 15$ visible.} \label{fig:r-300-small-nbbar}
\end{center}
\end{figure}

Here we investigate the case of $B \neq 0$ and, as an example, we choose $B=300$. In this case, obviously $\langle N_b \rangle _c = \langle \bar{N}_b \rangle _c + B$ and now 
$z_c = [\langle N_b \rangle_c (\langle N_b \rangle _c - B)]^{1/2}$. 
In general $\hat{R}^{(n,m)}$ is more complicated than for $B=0$ and only for very large $z_c$ or $\langle N_b \rangle _c$ it asymptotically approaches a linear function. This is demonstrated in Fig. \ref{fig:r-300}, where we plot $\hat{R}^{(n,m)}$ divided by $z_c$ as a function of $\langle \bar{N}_b\rangle_c$.
We were unable to obtain a simple approximated formula and thus in Fig. \ref{fig:r-300} we present only exact $\hat{R}^{(n,m)}/z_c$ based on \Crefrange{eq:c10}{eq:c33}. 
In the case of $B \neq 0$, $\hat{R}^{(n,0)} \neq \hat{R}^{(0,n)}$ and we decided to plot $(\hat{R}^{(n,0)} - (-1)^{n-1} (n-1)! \langle N_b \rangle _c)/z_c$ because this is symmetric when baryons and antibaryons are exchanged, see \Crefrange{eq:c10}{eq:c33}. 
We note that for some $\hat{R}^{(n,m)}/z_c$ with $n$, $m$ close to each other (e.g., $\hat{R}^{(2,2)}$, $\hat{R}^{(3,2)}$) we observe a maximum or minimum at $\langle \bar{N}_b\rangle_c$ about 100.
Experimentally available cases at heavy-ion colliders cover the values of $\langle \bar{N}_b \rangle _c$ of the order of 100 and in Fig. \ref{fig:r-300-small-nbbar} we show the results (except $\hat{R}^{(n,0)}$) in the range of $0 < \langle \bar{N}_b \rangle _c < 50$.  

\section{Comments and summary}

In this paper we calculated the proton, antiproton and mixed proton-antiproton factorial cumulants, $\hat{C}^{(n,m)}$, up to the sixth order, $n+m=6$, assuming that the only source of correlations is the global conservation of baryon number. The exact formulae are given in \Crefrange{eq:c10}{eq:c33} and for the case of $B=0$ the asymptotic expressions are provided in \Crefrange{eq:r20-of-zc}{eq:r33-of-zc}. The latter ones represent very good approximation already from $\langle N_b \rangle _c \approx 2$.

Several comments are in order.

Recently the ALICE Collaboration measured \cite{Acharya:2019izy} the second-order cumulant, $\kappa_2$, of the net-proton number and the result is consistent with the global baryon conservation. We note that, e.g., $\kappa_2$ contains less information than the second-order factorial cumulants $\hat{C}^{(2,0)}$, $\hat{C}^{(1,1)}$ and $\hat{C}^{(0,2)}$. It would be instructive to see whether the second-order factorial cumulants are consistent with the ALICE data. Also, the measurement of the higher-order factorial cumulants would be warranted. 

Having all the factorial cumulants we can immediately calculate the net-proton cumulants $\kappa_n$. 
For example \cite{Bzdak:2016sxg}
\begin{equation}
\kappa_{2}=\hat{C}^{(1,0)}+\hat{C}^{(0,1)}+\hat{C}^{(2,0)}+\hat{C}^{(0,2)}-2\hat{C}^{(1,1)} ,
\end{equation}
and the expressions for the higher order $\kappa_n$ are shown in Appendix \ref{appendix:appD}. Here $\hat{C}^{(1,0)}$ and $\hat{C}^{(0,1)}$ are the mean numbers of observed, e.g., protons and antiprotons, respectively. 

Finally, one possible way to measure factorial cumulants $\hat{C}^{(n,m)}$ is to first measure factorial moments $F_{i,k} \equiv \langle \frac{n!}{(n-i)!}\frac{\bar{n}!}{(\bar{n}-k)!}\rangle$,
which allow to directly obtain $\hat{C}^{(n,m)}$. Explicit relations between $\hat{C}^{(n,m)}$ and $F_{i,k}$ are given in Appendix \ref{appendix:appC}.

\begin{acknowledgements}
 This work was partially supported by the Ministry of Science and Higher Education, and by 
 the National Science Centre, Grant No. 2018/30/Q/ST2/00101.
\end{acknowledgements}


\appendix 
\section{A comment on Eq. (\ref{eq:distr2})} \label{appendix:appA}
Let in each heavy-ion collision event $B = N_{p}+N_{n}-\bar{N}_{p}-\bar{N}_{n}$ be the net-baryon number. Here $N_p$ and $\bar{N}_p$ are the total numbers of protons and antiprotons, respectively, $N_n$ and $\bar{N}_n$ are the total numbers of neutrons and antineutrons. Moreover, by $n_p$ and $\bar{n}_p$ we denote the numbers of observed protons and antiprotons in a given acceptance bin. $p_1=\langle n_p \rangle / \langle N_p \rangle$ is the probability to observe a proton in a given acceptance region and $p_2 =\langle \bar{n}_p \rangle / \langle \bar{N}_p \rangle$ is the probability to observe an antiproton.  
The probability distribution of $n_p$ and $\bar{n}_p$ is given by
\begin{equation}
\begin{aligned} \label{eq:distr1}
P({n}_{p},\bar{n}_{p})&=A \sum_{N_{p}={n}_{p}}^{\infty}\sum_{\bar{N}_{p}=\bar{n}_{p}}^{\infty}\sum_{N_{n}=0}^{\infty}\sum_{\bar{N}_{n}=0}^{\infty}\delta_{N_{p}+N_{n}-\bar{N}_{p}-\bar{N}_{n},B}\left[\frac{\langle N_{p}\rangle^{N_{p}}}{N_{p}!}e^{-\langle N_{p}\rangle}\right]\left[\frac{\langle \bar{N}_{p}\rangle^{\bar{N}_{p}}}{\bar{N}_{p}!}e^{-\langle \bar{N}_{p}\rangle}\right] \\ &\times \left[\frac{\langle N_{n}\rangle^{N_{n}}}{N_{n}!}e^{-\langle N_{n}\rangle}\right]\left[\frac{\langle \bar{N}_{n}\rangle^{\bar{N}_{n}}}{\bar{N}_{n}!}e^{-\langle \bar{N}_{n}\rangle}\right] \\ &\times \left[\frac{N_{p}!}{{n}_{p}!(N_{p}-{n}_{p})!}p_{1}^{{n}_{p}}(1-p_{1})^{N_{p}-{n}_{p}}\right]\left[\frac{\bar{N}_{p}!}{\bar{n}_{p}!(\bar{N}_{p}-\bar{n}_{p})!}p_{2}^{\bar{n}_{p}}(1-p_{2})^{\bar{N}_{p}-\bar{n}_{p}}\right],
\end{aligned}
\end{equation}
where $A$ is a normalization factor. In this expression we assume that the only source of correlation is given by the global conservation of baryon number implemented by $\delta_{N_{p}+N_{n}-\bar{N}_{p}-\bar{N}_{n},B}$.

Next, $N_{b} = N_p + N_n$ is the total number of baryons, and $\bar{N}_b = \bar{N}_p + \bar{N}_n$ is the total number of anti-baryons. Using relations
\begin{equation}
N_p=N_{b}-N_n, \quad \bar{N}_p=\bar{N}_b-\bar{N}_n\,,
\end{equation}
and summing over $N_n$ and $\bar{N}_n$ leads to our starting Eq. (\ref{eq:distr2}).

\section{Asymptotic expansion for $B=0$} 
\label{appendix:appE}

Here we present more details leading to the asymptotic \Crefrange{eq:r20-of-zc}{eq:r33-of-zc}. As already mentioned in Section \ref{sec:Beq0}, in all \Crefrange{eq:c10}{eq:c33} we eliminate the Bessel functions (the higher the order of the factorial cumulant, the more terms are needed in Eq. (\ref{eq:asympt-bessel}) and it is enough to take the first $7$ terms for the sixth order $\hat{C}^{(n,m)}$) and expand $\hat{R}^{(n,m)}(z)$ into a power series for large $z$. We obtain: 
{ 
\setlength{\abovedisplayskip}{6pt}
\setlength{\belowdisplayskip}{6pt}
\begin{fleqn}
\begin{gather}
 	\begin{aligned} \label{eq:r20-of-z}
 		\hat{R}^{(2,0)}(z) \sim -\tfrac{1}{2}z + \tfrac{1}{4} + \tfrac{3}{64}z^{-1} + ...
 	\end{aligned} \\
 	\begin{aligned}
 		\hat{R}^{(1,1)}(z) \sim \tfrac{1}{2}z + \tfrac{1}{64}z^{-1} + ...
 	\end{aligned}
\end{gather}
\begin{gather}
 	\begin{aligned}
 		\hat{R}^{(3,0)}(z) \sim \tfrac{3}{4}z - \tfrac{1}{2} - \tfrac{15}{128}z^{-1} + ...
 	\end{aligned} \\
 	\begin{aligned}
 		\hat{R}^{(2,1)}(z) \sim -\tfrac{1}{4}z - \tfrac{3}{128}z^{-1} + ...
 	\end{aligned} 
\end{gather}
\begin{gather}
 	\begin{aligned}
 		\hat{R}^{(4,0)}(z) \sim -\tfrac{15}{8}z + \tfrac{3}{2} + \tfrac{105}{256}z^{-1} + ...
 	\end{aligned} \\
 	\begin{aligned}
 		\hat{R}^{(3,1)}(z) \sim \tfrac{3}{8}z + \tfrac{15}{256}z^{-1} + ...
 	\end{aligned} \\
 	\begin{aligned}
 		\hat{R}^{(2,2)}(z) \sim \tfrac{1}{8}z + \tfrac{9}{256}z^{-1} + ...
 	\end{aligned} 
\end{gather}
\begin{gather}
 	\begin{aligned}
 		\hat{R}^{(5,0)}(z) \sim \tfrac{105}{16}z -6 - \tfrac{945}{512}z^{-1} + ...
 	\end{aligned} \\
 	\begin{aligned}
 		\hat{R}^{(4,1)}(z) \sim -\tfrac{15}{16}z - \tfrac{105}{512}z^{-1} + ...
 	\end{aligned} \\
 	\begin{aligned}
 		\hat{R}^{(3,2)}(z) \sim -\tfrac{3}{16}z - \tfrac{45}{512}z^{-1} + ...
 	\end{aligned} 
\end{gather}
\begin{gather}
 	\begin{aligned}
 		\hat{R}^{(6,0)}(z) \sim -\tfrac{945}{32}z +30 + \tfrac{10395}{1024}z^{-1} + ...
 	\end{aligned} \\
 	\begin{aligned}
 		\hat{R}^{(5,1)}(z) \sim \tfrac{105}{32}z + \tfrac{945}{1024}z^{-1} + ...
 	\end{aligned} \\
 	\begin{aligned}
 		\hat{R}^{(4,2)}(z) \sim \tfrac{15}{32}z + \tfrac{315}{1024}z^{-1} + ...
 	\end{aligned} \\
 	\begin{aligned}\label{eq:r33-of-z}
 		\hat{R}^{(3,3)}(z) \sim \tfrac{9}{32}z + \tfrac{225}{1024}z^{-1} + ...
 	\end{aligned} 
\end{gather}
\end{fleqn}
}

Note that all the $\hat{R}^{(n,m)}(z)$ can be written as
\begin{equation}\label{eq:r-gen-of-z}
\hat{R}^{(n,m)}(z) \sim a_1 z + a_0 + a_{-1} z^{-1} + a_{-2} z^{-2} + ...\,,
\end{equation}
where the coefficients $a_i$ depend on $n$ and $m$ and $a_0 \neq 0$ for $m=0$ only.

It is easy to see that $\hat{R}^{(n,m)}(z_c)$ is also of the same form, that is, the highest term is proportional to $z_c$ and the coefficients of the series can be easily calculated. 
First, let us expand $z_c$ in a series of $z$:
\begin{equation} \label{eq:zc-of-z}
z_c(z) \sim z - \frac{1}{4} - \frac{1}{32}z^{-1} - \frac{1}{64}z^{-2}  ...
\end{equation} 

It is clear that $\hat{R}^{(n,m)}(z_c)$ cannot have a $z_{c}^2$ term (or higher order) because it would generate a $z^2$ term in $\hat{R}^{(n,m)}(z)$ and we know that this term is not present, see Eq. (\ref{eq:r-gen-of-z}). Thus $\hat{R}^{(n,m)}(z_c)$ can be written as  
\begin{equation} \label{eq:r-of-zc-test-2}
\hat{R}^{(n,m)}(z_c) \sim b_1 z_c + b_0 + b_{-1} z_c^{-1} + b_{-2} z_c^{-2} + ...\,,
\end{equation}
where the coefficients $b_i$ are to be determined. Substituting Eq. (\ref{eq:zc-of-z}) into Eq. (\ref{eq:r-of-zc-test-2}) and comapring with Eq. (\ref{eq:r-gen-of-z}) we obtain: 
\begin{fleqn}
\begin{gather}
 	\begin{aligned} 
 		b_1=a_1 \,,
 	\end{aligned} \\
 	\begin{aligned}
 		b_0=a_0+ \tfrac{1}{4}a_1 \,,
 	\end{aligned}\\
 	\begin{aligned}
 		b_{-1}=a_{-1} + \tfrac{1}{32}a_1 \,,
 	\end{aligned} \\
 	\begin{aligned}
 		b_{-2}=a_{-2} - \tfrac{1}{4}a_{-1} + \tfrac{1}{128}a_1\,.
 	\end{aligned} 
\end{gather}
\end{fleqn}

Clearly, this procedure may be easily extended to obtain more terms if needed. These relations combined with \Crefrange{eq:r20-of-z}{eq:r33-of-z} lead to our \Crefrange{eq:r20-of-zc}{eq:r33-of-zc}.

\section{Net-proton cumulants} 
\label{appendix:appD}
The cumulant generating function for two species of particles reads
\begin{equation} \label{eq:cgf}
K(t, \bar{t})= G\left(e^t, e^{\bar{t}}\right)\,,
\end{equation}
where $G(x, \bar{x})$ is given by Eq. (\ref{eq:fcgf}). In particular, the net-particle (e.g. net-proton) cumulants are given by ($\bar{t}=-t$)
\begin{equation} \label{eq:cumulants}
\kappa_i = \left. \frac{d^i}{dt^i}K(t, -t) \right\rvert_{t=0}\,.
\end{equation}
Combining Eqs. (\ref{eq:cgf}) and (\ref{eq:cumulants}), we have:
\begin{equation} 
\kappa_i = \left. \frac{d^i}{dt^i}G(x(t), \bar{x}(t)) \right\rvert_{t=0}\,,
\end{equation}
where $x(t) = e^t$ and $\bar{x}(t)=e^{-t}$ and hence derivatives $x^{(n)}(t{=}0)=1$, $\bar{x}^{(n)}(t{=}0)=(-1)^n$. Using this and Eq. (\ref{eq:fact-cum}), we obtain the formulae for the net-proton cumulants in terms of the factorial cumulants: 
\begin{fleqn} 
\begin{gather}
\begin{aligned}
	\kappa_{1}=\hat{C}^{(1,0)}-\hat{C}^{(0,1)} \,,
\end{aligned} \\
\begin{aligned}
	\kappa_{2}=\hat{C}^{(1,0)}+\hat{C}^{(0,1)}+\hat{C}^{(2,0)}+\hat{C}^{(0,2)}-2\hat{C}^{(1,1)} \,,
\end{aligned} \\
\begin{aligned}
	\kappa_{3}=\hat{C}^{(1,0)}-\hat{C}^{(0,1)}+3\left(\hat{C}^{(2,0)}-\hat{C}^{(0,2)}\right)+\hat{C}^{(3,0)}-\hat{C}^{(0,3)}-3\left(\hat{C}^{(2,1)}-\hat{C}^{(1,2)}\right) \,,
\end{aligned} \\
\begin{aligned}
	\kappa_{4}&=\hat{C}^{(1,0)}+\hat{C}^{(0,1)}+7\left(\hat{C}^{(2,0)}+\hat{C}^{(0,2)}\right)-2\hat{C}^{(1,1)}+6\left(\hat{C}^{(3,0)}+\hat{C}^{(0,3)}\right)-6\left(\hat{C}^{(2,1)}+\hat{C}^{(1,2)}\right)\\&+\hat{C}^{(4,0)}+\hat{C}^{(0,4)}-4\left(\hat{C}^{(3,1)}+\hat{C}^{(1,3)}\right)+6\hat{C}^{(2,2)} \,,
\end{aligned} \\
\begin{aligned}
	\kappa_{5}&=\hat{C}^{(1,0)}-\hat{C}^{(0,1)}+15\left(\hat{C}^{(2,0)}-\hat{C}^{(0,2)}\right)+25\left(\hat{C}^{(3,0)}-\hat{C}^{(0,3)}\right)-15\left(\hat{C}^{(2,1)}-\hat{C}^{(1,2)}\right)\\&+10\left(\hat{C}^{(4,0)}-\hat{C}^{(0,4)}\right)-20\left(\hat{C}^{(3,1)}-\hat{C}^{(1,3)}\right)+\hat{C}^{(5,0)}-\hat{C}^{(0,5)}-5\left(\hat{C}^{(4,1)}-\hat{C}^{(1,4)}\right)\\&+10\left(\hat{C}^{(3,2)}-\hat{C}^{(2,3)}\right) \,,
\end{aligned} \\
\begin{aligned}
	\kappa_{6}&=\hat{C}^{(1,0)}+\hat{C}^{(0,1)}+31\left(\hat{C}^{(2,0)}+\hat{C}^{(0,2)}\right)-2\hat{C}^{(1,1)}+90\left(\hat{C}^{(3,0)}+\hat{C}^{(0,3)}\right)\\&-30\left(\hat{C}^{(2,1)}+\hat{C}^{(1,2)}\right)+65\left(\hat{C}^{(4,0)}+\hat{C}^{(0,4)}\right)-80\left(\hat{C}^{(3,1)}+\hat{C}^{(1,3)}\right)+30\hat{C}^{(2,2)}\\&+15\left(\hat{C}^{(5,0)}+\hat{C}^{(0,5)}\right)-45\left(\hat{C}^{(4,1)}+\hat{C}^{(1,4)}\right)+30\left(\hat{C}^{(3,2)}+\hat{C}^{(2,3)}\right)+\hat{C}^{(6,0)}+\hat{C}^{(0,6)}\\&-6\left(\hat{C}^{(5,1)}+\hat{C}^{(1,5)}\right)+15\left(\hat{C}^{(4,2)}+\hat{C}^{(2,4)}\right)-20\hat{C}^{(3,3)} \,,
\end{aligned}
\end{gather}
\end{fleqn}
where $\hat{C}^{(1,0)}$ and $\hat{C}^{(0,1)}$ are the mean numbers of, e.g., protons and antiprotons, respectively. These results extend the formulae provided in Appendix A of Ref. \cite{Bzdak:2016sxg}.

\section{$\hat{C}^{(n,m)}$ vs $F_{i,k}$} 
\label{appendix:appC}
The factorial moments for two variables (two species of particles) are defined via the factorial moment generating function $H(x,\bar{x})$ (see Eq. (\ref{eq:hfun})):
\begin{equation}
F_{i,k}\equiv \left\langle \frac{n_{1}!}{(n_{1}-i)!}\frac{n_{2}!}{(n_{2}-k)!} \right\rangle = \left. \frac{d^{i}}{dx^{i}}\frac{d^{k}}{d\bar{x}^{k}}H(x,\bar{x}) \right\rvert_{x=\bar{x}=1}.
\end{equation}
Using Eqs. (\ref{eq:fcgf}) and (\ref{eq:fact-cum}), and the normalization condition $H(1,1)=1$, we can express the factorial cumulants through the factorial moments:
{ 
\setlength{\abovedisplayskip}{6pt}
\setlength{\belowdisplayskip}{6pt}
\begin{fleqn} 
\begin{gather}
\begin{aligned}
	\hat{C}^{(1,0)}&=F_{1,0}
\end{aligned} \\
\begin{aligned} 
	\hat{C}^{(0,1)}&=F_{0,1} 
\end{aligned}
\end{gather}
\begin{gather}
\begin{aligned} \hat{C}^{(2,0)}=-F_{1,0}^{2}+F_{2,0} \end{aligned} \\
\begin{aligned} \hat{C}^{(1,1)}=-F_{0,1}F_{1,0}+F_{1,1} \end{aligned}
\end{gather}
\begin{gather}
\begin{aligned} \hat{C}^{(3,0)}=2F_{1,0}^{3}-3F_{1,0}F_{2,0}+F_{3,0} \end{aligned} \\
\begin{aligned}
\hat{C}^{(2,1)}&=2F_{0,1}F_{1,0}^{2}-2F_{1,0}F_{1,1}-F_{0,1}F_{2,0}+F_{2,1}
\end{aligned}
\end{gather}
\begin{gather}
\begin{aligned}
\hat{C}^{(4,0)}&=-6F_{1,0}^{4}+12F_{1,0}^{2}F_{2,0}-3F_{2,0}^{2}-4F_{1,0}F_{3,0}+F_{4,0}
\end{aligned} \\
\begin{aligned}
\hat{C}^{(3,1)}&=-6F_{0,1}F_{1,0}^{3}+6F_{1,0}^{2}F_{1,1}+6F_{0,1}F_{1,0}F_{2,0}-3F_{1,1}F_{2,0}-3F_{1,0}F_{2,1}-F_{0,1}F_{3,0}+F_{3,1}
\end{aligned} \\
\begin{aligned}
\hat{C}^{(2,2)}&=\left(-6F_{0,1}^{2}+2F_{0,2}\right)F_{1,0}^{2}+8F_{0,1}F_{1,0}F_{1,1}-2F_{1,1}^{2}-2F_{1,0}F_{1,2}+\left(2F_{0,1}^{2}-F_{0,2}\right)F_{2,0}\\&-2F_{0,1}F_{2,1}+F_{2,2}
\end{aligned}
\end{gather}
\begin{gather}
\begin{aligned}
\hat{C}^{(5,0)}&=24F_{1,0}^{5}-60F_{1,0}^{3}F_{2,0}+30F_{1,0}F_{2,0}^{2}+20F_{1,0}^{2}F_{3,0}-10F_{2,0}F_{3,0}-5F_{1,0}F_{4,0}+F_{5,0}
\end{aligned} \\
\begin{aligned}
\hat{C}^{(4,1)}&=24F_{0,1}F_{1,0}^{4}-24F_{1,0}^{3}F_{1,1}-36F_{0,1}F_{1,0}^{2}F_{2,0}+24F_{1,0}F_{1,1}F_{2,0}+6F_{0,1}F_{2,0}^{2}+12F_{1,0}^{2}F_{2,1}\\&-6F_{2,0}F_{2,1}+8F_{0,1}F_{1,0}F_{3,0}-4F_{1,1}F_{3,0}-4F_{1,0}F_{3,1}-F_{0,1}F_{4,0}+F_{4,1}
\end{aligned} \\
\begin{aligned}
\hat{C}^{(3,2)}&=2\left(12F_{0,1}^{2}-3F_{0,2}\right)F_{1,0}^{3}-36F_{0,1}F_{1,0}^{2}F_{1,1}+12F_{1,0}F_{1,1}^{2}+6F_{1,0}^{2}F_{1,2}\\&-3\left(6F_{0,1}^{2}-2F_{0,2}\right)F_{1,0}F_{2,0}+12\left(F_{1,1}F_{2,0}+F_{1,0}F_{2,1}\right)F_{0,1}-3F_{1,2}F_{2,0}-6F_{1,1}F_{2,1}\\&-3F_{1,0}F_{2,2}+\left(2F_{0,1}^{2}-F_{0,2}\right)F_{3,0}-2F_{0,1}F_{3,1}+F_{3,2}
\end{aligned}
\end{gather}
\begin{gather}
\begin{aligned}
\hat{C}^{(6,0)}&=-120 F_{1,0}^6+360 F_{2,0}
   F_{1,0}^4-120 F_{3,0} F_{1,0}^3-270 F_{2,0}^2
   F_{1,0}^2+30 F_{4,0} F_{1,0}^2+120 F_{2,0}
   F_{3,0} F_{1,0}\\&-6 F_{5,0} F_{1,0}+30
   F_{2,0}^3-10 F_{3,0}^2-15 F_{2,0}
   F_{4,0}+F_{6,0}
\end{aligned} \\
\begin{aligned}
\hat{C}^{(5,1)}&=-120 F_{0,1} F_{1,0}^5+120 F_{1,1} F_{1,0}^4+240
   F_{0,1} F_{2,0} F_{1,0}^3-60 F_{2,1}
   F_{1,0}^3-180 F_{1,1} F_{2,0} F_{1,0}^2\\&-60
   F_{0,1} F_{3,0} F_{1,0}^2+20 F_{3,1}
   F_{1,0}^2-90 F_{0,1} F_{2,0}^2 F_{1,0}+60
   F_{2,0} F_{2,1} F_{1,0}+40 F_{1,1} F_{3,0}
   F_{1,0}\\&+10 F_{0,1} F_{4,0} F_{1,0}-5 F_{4,1}
   F_{1,0}+30 F_{1,1} F_{2,0}^2+20 F_{0,1}
   F_{2,0} F_{3,0}-10 F_{2,1} F_{3,0}-10 F_{2,0}
   F_{3,1}\\&-5 F_{1,1} F_{4,0}-F_{0,1}
   F_{5,0}+F_{5,1}
\end{aligned} \\
\begin{aligned}
\hat{C}^{(4,2)}&=-6 \left(20 F_{0,1}^2-4 F_{0,2}\right)
   F_{1,0}^4+192 F_{0,1} F_{1,1} F_{1,0}^3+12
   \left(12 F_{0,1}^2-3 F_{0,2}\right) F_{2,0}
   F_{1,0}^2\\&-4 \left(6 F_{0,1}^2-2
   F_{0,2}\right) F_{3,0} F_{1,0}-3 \left(6
   F_{0,1}^2-2 F_{0,2}\right) F_{2,0}^2-6
   \left(4 F_{1,2} F_{1,0}^3+12 F_{1,1}^2
   F_{1,0}^2\right)\\&+24 F_{0,1} F_{2,0}
   F_{2,1}-72 F_{0,1} \left(F_{2,1} F_{1,0}^2+2
   F_{1,1} F_{2,0} F_{1,0}\right)\\&+12
   \left(F_{2,2} F_{1,0}^2+4 F_{1,1} F_{2,1}
   F_{1,0}+\left(2 F_{1,1}^2+2 F_{1,0}
   F_{1,2}\right) F_{2,0}\right)-3 \left(2
   F_{2,1}^2+2 F_{2,0} F_{2,2}\right)\\&+16 F_{0,1}
   \left(F_{1,1} F_{3,0}+F_{1,0}
   F_{3,1}\right)-4 \left(F_{1,2} F_{3,0}+2
   F_{1,1} F_{3,1}+F_{1,0}
   F_{3,2}\right)\\&+\left(2
   F_{0,1}^2-F_{0,2}\right) F_{4,0}-2 F_{0,1}
   F_{4,1}+F_{4,2}
\end{aligned} \\
\begin{aligned}
\hat{C}^{(3,3)}&=2 \left(-60 F_{0,1}^3+36 F_{0,2} F_{0,1}-3
   F_{0,3}\right) F_{1,0}^3+18 \left(12
   F_{0,1}^2-3 F_{0,2}\right) F_{1,1}
   F_{1,0}^2\\&-3 \left(-24 F_{0,1}^3+18 F_{0,2}
   F_{0,1}-2 F_{0,3}\right) F_{2,0} F_{1,0}-18
   F_{0,1} \left(3 F_{1,2} F_{1,0}^2+6 F_{1,1}^2
   F_{1,0}\right)\\&+2 \left(6 F_{1,1}^3+18 F_{1,0}
   F_{1,2} F_{1,1}+3 F_{1,0}^2 F_{1,3}\right)-9
   \left(6 F_{0,1}^2-2 F_{0,2}\right)
   \left(F_{1,1} F_{2,0}+F_{1,0}
   F_{2,1}\right)\\&+18 F_{0,1} \left(F_{1,2}
   F_{2,0}+2 F_{1,1} F_{2,1}+F_{1,0}
   F_{2,2}\right)\\&-3 \left(F_{1,3} F_{2,0}+3
   F_{1,2} F_{2,1}+3 F_{1,1} F_{2,2}+F_{1,0}
   F_{2,3}\right)+\left(-6 F_{0,1}^3+6 F_{0,2}
   F_{0,1}-F_{0,3}\right) F_{3,0}\\&+3 \left(2
   F_{0,1}^2-F_{0,2}\right) F_{3,1}-3 F_{0,1}
   F_{3,2}+F_{3,3}.
   \end{aligned}
\end{gather}
\end{fleqn}
}
These results extend the formulae provided in Appendix A of Ref. \cite{Bzdak:2016sxg}.

\bibliography{paper}

\end{document}